\documentclass[a4paper,fleqn,usenatbib]{mnras}
\usepackage[T1]{fontenc}
\usepackage{ae,aecompl}
\usepackage{graphicx}
\usepackage{amsmath}
\usepackage{amssymb}
\usepackage{txfonts}

\def\spirou#1{{#1}}

\newcommand{\ltsima}{$\; \buildrel < \over \sim \;$}
\newcommand{\lsim}{\lower.5ex\hbox{\ltsima}}
\newcommand{\gtsima}{$\; \buildrel > \over \sim \;$}
\newcommand{\gsim}{\lower.5ex\hbox{\gtsima}}
\newcommand{\bra}{\langle}
\newcommand{\ket}{\rangle}

\newcommand{\dd}{\mathrm{d}}

\newcommand{\ci}{\mathrm{i}}

\newcommand{\vecl}{\bmath{\ell}}
\newcommand{\veck}{\bmath{k}}
\newcommand{\veclp}{\bmath{\ell}^\prime}
\newcommand{\veckp}{\bmath{k^\prime}}
\newcommand{\trace}{\mathrm{tr}}

\newcommand{\vecdlp}{\vecl-\veclp}
\newcommand{\vecdkp}{\veck-\veckp}

\newcommand{\chip}{{\chi^\prime}}

\newcommand{\lprime}{\ell^\prime}

\newcommand{\kp}{k^\prime}

\title[Second order effects to weak cosmic shear]
{Gravitational corrections to light propagation in a perturbed FLRW-universe and corresponding weak lensing spectra}
\author[C. Cuesta-Lazaro, A. Quera-Bofarull, R. Reischke, B.M. Sch{\"a}fer]
{Carolina Cuesta-Lazaro, Arnau Quera-Bofarull, Robert Reischke, Bj{\"o}rn Malte Sch{\"a}fer\thanks{e-mail: bjoern.malte.schaefer@uni-heidelberg.de}\\
Astronomisches Rechen-Institut, Zentrum f{\"u}r Astronomie der Universit{\"a}t Heidelberg, Philosophenweg 12, 69120 Heidelberg, Germany}

\begin{document}
\onecolumn
\pagerange{\pageref{firstpage}--\pageref{lastpage}}
\pubyear{2017}
\maketitle
\label{firstpage}

\begin{abstract}
When the gravitational lensing of the large-scale structure is calculated from a cosmological model a few assumptions enter: $(i)$ one assumes that the photons follow unperturbed background geodesics, which is usually referred to as the Born-approximation, $(ii)$ the lenses move slowly, $(iii)$ the source-redshift distribution is evaluated relative to the background quantities and $(iv)$ the lensing effect is linear in the gravitational potential. Even though these approximations are small individually they could sum up, especially since they include local effects such as the Sachs-Wolfe and peculiar motion, but also non-local ones like the Born-approximation and the integrated Sachs-Wolfe effect. In this work we will address all points mentioned and perturbatively calculate the effect on a tomographic cosmic shear power spectrum of each effect individually as well as all cross-correlations. Our findings show that each effect is at least 4 to 5 orders of magnitude below the leading order lensing signal. Finally we sum up all effects to estimate the overall impact on parameter estimation by a future cosmological weak lensing survey such as \textit{Euclid} in a $w$CDM cosmology with parametrisation $\Omega_\mathrm{m}$, $\sigma_8$, $n_\mathrm{s}$, $h$, $w_0$ and $w_\mathrm{a}$, using 5 tomographic bins. We consistently find a parameter bias of $10^{-5}$, which is therefore completely negligible for all practical purposes, confirming that other effects such as intrinsic alignments and magnification bias will be the dominant systematic source in future surveys.
\end{abstract}


\section{Introduction}
Measuring weak gravitational lensing of the large-scale structure \cite[LSS,][]{2000astro.ph..3338K,2000MNRAS.318..625B,VanWaerbeke2000,2007MNRAS.381.1018A,hoekstra_weak_2008,kilbinger_dark-energy_2009,
kayo_information_2013,kitching_3d_2014,Kilbinger2015}, called cosmic shear, is a key scientific goal in upcoming cosmological surveys such as \textit{Euclid} \citep{EuclidStudyReport} or the \textit{Large Synoptic Survey Telescope} \citep[LSST][]{Collaboration2009}. These surveys will provide data over a vast range of scales. In particular they will probe the small scales with unrivalled precision allowing to measure the cosmic shear signal at roughly 1000$\sigma$ significance. Measuring the weak lensing signal at such a high significance is a challenge on its own \citep[see e.g.][]{Briddle2008,Miller2013,Croft2017,Hoekstra2017}, but also on the theoretical side quite a few systematics exists, most notably intrinsic alignments of galaxies, which can mimic a weak lensing signal \citep[e.g.][]{croft_weak-lensing_2000, lee_intrinsic_2011, joachimi_galaxy_2015, schaefer_angular_2015, blazek_tidal_2015}. 

To infer, in the very end, cosmological parameters from cosmic shear measurements the lensing signal must be calculated from a cosmological model and include all effects present in the observation. Otherwise systematic effects in the inference process can occur \citep{amara_systematic_2008}. The cosmic shear signal is usually given in tomographic bins to restore redshift information \citep{1999ApJ...522L..21H,2004ApJ...601L...1T,2007MNRAS.381.1018A} or in a spherical basis using spherical Bessel functions \citep{2003MNRAS.343.1327H,2005PhRvD..72b3516C}. Furthermore the lensing potential is calculated in a weakly perturbed spacetime relative to a Friedmann-Lema{\^i}tre-Robertson-Walker by tracing null geodesics by virtue of the Jacobi equation, describing how a bundle of light rays is deformed along their path. The equation of geodesic deviation is then sourced by gradients in the Newtonian potential, $\Phi$, or in presence of anisotropic stress by the difference of the two Bardeen Potentials. Since $\Phi \ll c^2$, deflection angles are small and the integration is usually carried out along unperturbed rays, this is known as the Born-approximation, which is satisfied very well as shown by different numerical simulations \citep{2000ApJ...530..547J, Dodelson2005, Shapiro:2006em, Hilbert2009}. A second order treatment can also be found in \citet{2002ApJ...574...19C, PhysRevD.81.083002}. Peculiar velocity effects have been discussed in \citet{Bonvin2008}, while other systematic effects such as multiple deflections have been treated in \citep{1994A&A...287..349S, Krause:2009yr}.

The statistical properties of the lensing potential of the large scale structures are then expressed in terms of angular correlation function in a suitable basis \citep{1999ApJ...522L..21H,2003MNRAS.343.1327H,2012MNRAS.423.3445S,Kitching2017}. At its heart it involves a line of sight projection of the power spectrum of the potential fluctuations, or equivalently of the matter power spectrum. Cosmological information is mainly contained in the lensing kernel and especially in the distance redshift relation. As described before, the natural coordinate choice for weak cosmological lensing are comoving coordinates together with conformal time, however  only the redshift $z$ is observable, which has to be related to the comoving distance $\chi$. In calculating the cosmic shear power spectrum one thus assumes sources measured at redshift $z$ to be at unperturbed comoving distances $\chi$. However redshift space distortions \citep[e.g.][]{Percival2011} add an additional contribution to the cosmological redshift. The observed redshift distributions does therefore not correspond to the ideally assumed one used for the calculation of the cosmic shear power spectrum. Furthermore, the Born-approximation results into a temporal and spatial part. The effect of the temporal Born-correction changes the evolutionary state at which a photon passes cosmic structures and therefore the lensing signal is changed. Other effects involve the kinetic contributions of the LSS to the lensing potential. In this paper we calculate the magnitude of those effects, particularly we investigate redshift space distortions, second-order correction to the effective refraction index, the temporal Born-approximation, Sachs-Wolfe effects and gravitomagnetic effects. We calculate the corrections to the tomographic cosmic shear signal by calculation the contribution of the auto- and cross-correlations of the effects mentioned. The effects are compared to weak lensing spectra as being measured by \textit{Euclid}.

Throughout the paper we will use a spatially flat $w$CDM-cosmology, with specific parameter choices $\Omega_m = 0.25$, $n_s = 1$, $\sigma_8 = 0.8$ and $h=0.7$ and $w=-1$ for the fiducial cosmology.
The structure of the paper is as follows: After a summary of cosmology in \autoref{sect_cosmology} and weak gravitational lensing in \autoref{sect_lensing} we work out all corrections in \autoref{sect_second} and evaluate them numerically in \autoref{sect_corrections}. We summarise and discuss our results in \autoref{sect_summary}.

\section{cosmology}\label{sect_cosmology}
Under the symmetry assumption of Friedmann-Lema{\^i}tre-cosmologies all fluids are characterised by their density and their equation of state: In spatially flat cosmologies with the matter density parameter $\Omega_\mathrm{m}$ and the corresponding dark energy density $1-\Omega_\mathrm{m}$ one obtains for the Hubble function $H(a)=\dot{a}/a$ the expression,
\begin{equation}
\frac{H^2(a)}{H_0^2} = \frac{\Omega_\mathrm{m}}{a^{3}} + \frac{1-\Omega_\mathrm{m}}{a^{3(1+w)}},
\end{equation}
for a constant dark energy equation of state parameter $w$. The comoving distance $\chi$ is related to the scale factor $a$ through
\begin{equation}
\chi = -c\int_1^a\:\frac{\dd a}{a^2 H(a)},
\end{equation}
where the Hubble distance $\chi_H=c/H_0$ sets the distance scale for cosmological distance measures. Cosmic deceleration $q=\ddot{a}a/\dot{a}^2$ is related to the logarithmic derivative of the Hubble function, $2-q = 3+\dd\ln H/\dd\ln a$.

Small fluctuations $\delta$ in the distribution of dark matter grow, as long as they are in the linear regime $\left|\delta\right|\ll 1$, according to the growth function $D_+(a)$ \citep{2003MNRAS.346..573L,1998ApJ...508..483W},
\begin{equation}
\frac{\dd^2}{\dd a^2}D_+(a) +
\frac{2-q}{a}\frac{\dd}{\dd a}D_+(a) -
\frac{3}{2a^2}\Omega_\mathrm{m}(a) D_+(a) = 0,
\label{eqn_growth}
\end{equation}
and their statistics is characterised by the spectrum $\bra \delta(\bmath{k})\delta^*(\bmath{k}^\prime)\ket = (2\pi)^3\delta_D(\bmath{k}-\bmath{k}^\prime)P_\delta(k)$. Inflation generates a spectrum of the form $P_\delta(k)\propto k^{n_s}T^2(k)$ with the transfer function $T(k)$ \citep{eisenstein_power_1999, Eisenstein1998} which is normalised to the variance $\sigma_8^2$ smoothed to the scale of $8~\mathrm{Mpc}/h$,
\begin{equation}
\sigma_8^2 = \int_0^\infty\frac{k^2\dd k}{2\pi^2}\: W^2(8~\mathrm{Mpc}/h\times k)\:P_\delta(k),
\end{equation}
with a Fourier-transformed spherical top-hat $W(x) = 3j_1(x)/x$ as the filter function. From the CDM-spectrum of the density perturbation the spectrum of the Newtonian gravitational potential $\Phi $ can be obtained,
\begin{equation}
	\label{eq::poisson_power}
	P_\Phi(k; \chi) = \left(\frac{3 \Omega_\mathrm{m} H_0^2}{2} \right)^2 k^{-4} P_\delta (k; \chi)   \propto \left(\frac{3\Omega_\mathrm{m}H_0^2}{2}\right)^2\:k^{n_s-4}\:T(k)^2,
\end{equation}
by applying the comoving Poisson equation $\Delta\Phi= \frac{3 \Omega_\mathrm{m}H_0^2}{2} \delta$  for deriving the gravitational potential $\Phi$ from the density $\delta$. With Eq. (\ref{eqn_growth}) yielding a solution for the homogeneous growth of the density contrast. It should be noted that velocities at linear order are obtained from the continuity equation,
\begin{equation}
	\boldsymbol{\nabla} \cdot \boldsymbol{\upsilon} = - a \dot{ \delta },
\end{equation}
such that in Fourier space,
\begin{equation}
	\boldsymbol{\upsilon}_k = a H(a) \frac{\dd \ln( D_+)}{\dd \ln(a)} \frac{ \boldsymbol{k}}{k^2} \boldsymbol{ \delta }_k.
\end{equation}

\section{Basics of cosmological weak gravitational lensing}\label{sect_lensing}
In weak gravitational lensing one investigates the action of gravitational tidal fields on the shape of distant galaxies by the distortion of light bundles \citep[for reviews, please refer to][]{2001PhR...340..291B, hoekstra_weak_2008, huterer_weak_2010, 2010CQGra..27w3001B,Kilbinger2015}.

The lensing potential $\psi$ is given by a projection integral,
\begin{equation}
  \psi(\boldsymbol{n}, \chi)
  = \frac{2}{c^2}\int_0^{\chi}\dd\chi'\:g(\chi,\chi')\Phi(\boldsymbol{n} \chi' ; \chi'),
\label{eqn_lensing_potential}
\end{equation}
relating $\psi$ to the gravitational potential $\Phi$ through weighting
function $g(\chi, \chi')$,
\begin{equation}
  g(\chi, \chi')= \frac{\chi'-\chi}{\chi'\chi}
\end{equation}
with $\boldsymbol{n}$ representing the position of the lens on the sky and $\chi$ its comoving
distance.
Since the intrinsic ellipticities of the galaxies are unknown, but assumed to be randomly ordered, it is convenient to average the lensing potential over a source distribution $p( z )$.
\begin{equation}
	\overline{\psi}( \boldsymbol{n} ) = 
	\int_0^{\chi_H} \dd\mathrm{\chi}\: p(z) \frac{\dd z}{\dd \chi} \psi ( \boldsymbol{n} , \chi ) = 
	\frac{2}{c^2} \int_0^{\chi_H} \dd\mathrm{\chi}\: \int_{\chi}^{\chi_H} \dd \chi' g( \chi , \chi') \Phi(\boldsymbol{n}\chi ; \chi)\; ,
\end{equation}
where we included the probability distribution inside the window function,
\begin{equation}
	g ( \chi , \chi') = \frac{ \chi' - \chi}{ \chi \chi} p( \chi') \frac{\!\dd z}{\dd \chi '}\;,
\end{equation}
and readjusted the integration boundaries. Note that $\dd z/\dd\chi' = H(a( \chi'))$.
As a line of sight-integrated quantity, the projected potential contains less information than the sourcing field $\Phi$. In order to partially regain that information, one commonly divides the sample of lensed galaxies into $n_\mathrm{bin}$ redshift bins and computes the lensing signal for each of the bins $i$ separately. Denoting $g_i ( \chi , \chi')$ as the restriction of $g( \chi , \chi')$ onto the bin $i$, one defines the tomographic lensing efficiency function $G_i(\chi)$,
\begin{equation}
	G_i(\chi) = \int^{ \chi }_{ 0}\dd\chip g_i( \chi, \chi') \:.
\end{equation}
Euclid forecasts use the parametrisation of the redshift distribution $p(z)\dd z$,
\begin{equation}
p(z)\dd z \propto \left(\frac{z}{z_0}\right)^2\exp\left[-\left(\frac{z}{z_0}\right)^\beta\right]\dd z,
\end{equation}
\spirou{with $\beta=3/2$ causing a slightly faster than exponential decrease at large redshifts \citep{EuclidStudyReport}}.

Changes in the image of a distant galaxy are encoded in the second angular derivatives of the weak lensing potential $\psi$: This Jacobian matrix can be decomposed into convergence and shear with the use of Pauli-matrices $\sigma_\alpha$,
\begin{equation}
\psi_{ab} 
= \sum_{\alpha=0}^3\: a_\alpha\sigma^{(\alpha)}
= \kappa\sigma^{(0)}_{ab} + \gamma_+\sigma^{(1)}_{ab} - \ci\rho\sigma^{(2)}_{ab} + \gamma_\times\sigma^{(3)}_{ab}\; .
\end{equation}
Since weak gravitational lensing of a single galaxy is not observable one is interested in the statistical properties of the convergence or the shear. The Fourier transform of the tomographic convergence correlation function, the angular power spectrum, is given by
\begin{equation}\label{eq:Ckappa_ij}
C^\kappa_{ij}(\ell) = 
\frac{9\Omega_\mathrm{m}^2}{16\chi_H^4}\int \frac{\mathrm{d}\chi}{\chi^2}\:g_i(\chi)g_j(\chi)P_\delta(\ell/\chi,\chi)\; ,
\end{equation}
where we used the Limber approximation \citep{1954ApJ...119..655L} and the comoving Poisson equation \eqref{eq::poisson_power}. Note that the spectrum \eqref{eq:Ckappa_ij} is equal to the $E$-mode spectrum of the weak lensing shear $\gamma$.

\section{Corrections to the weak lensing signal}\label{sect_second}
In this section we will describe the effects which will lead to corrections of the weak lensing signal. Particularly we are looking at the effects of peculiar velocity induced redshifts, Sachs-Wolfe effects, second order corrections to the effective speed of light, the temporal Born-approximation as well as the assumption of slowly moving lenses. All corrections originate from the metric perturbations $\Phi$, i.e. the Newtonian gravitational potential. The calculations are effectively carried out in synchronous Newtonian gauge.

\subsection{Distortions of the source redshift distribution}
Distance in galaxy surveys are measured indirectly via spectroscopic \citep{Gaztanaga2012,Cunha2014} or photometric \citep{Bolzonella2000,Bendera} redshift determinations. Mostly a combination of both techniques is used, such that the photometric method is calibrated with the spectroscopic one. When calculating the theoretical prediction of the lensing signal in Eq. (\ref{eq:Ckappa_ij}) one implicitly assumes that lensing takes places in the ideal background cosmology, expressed by the conversion of the redshift in a comoving distance. The perturbation on this background however alter the ideal cosmological redshift and the redshift distribution gets effectively distorted due to the presence of inhomogeneities. Consequently the lensing signal will look different as in (\ref{eq:Ckappa_ij}).

Quite generically, the kernel of the lensing potential can be expanded around its homogeneous value (see Eq \ref{eqn_lensing_potential})
\begin{equation}
	\psi _{ab} (\boldsymbol{n}, \chi + \Delta \chi) \approx \psi_{ab}(\boldsymbol{n}, \chi) + \frac{\partial \psi_{ab}(\boldsymbol{n}, \chi)}{\partial \chi} \Delta \chi = \psi_{ab}(\boldsymbol{n}, \chi)
+ \frac{\partial \psi_{ab}(\boldsymbol{n}, \chi)}{\partial \chi} \frac{\partial \chi}{\partial z}\Delta z\; .
\end{equation}
Here we keep the indices $a,b$ as bookkeeping for the derivatives. We now rewrite the latter equation by using the Leibniz rule,
\begin{equation}
	\psi_{ab}(\boldsymbol{n}, \chi) = \frac{2}{c^2 } \int_0^\chi \mathrm{d} \chip g(\chip, \chi) S_{ab}((\boldsymbol{n} \chi; \chi) ,(\boldsymbol{n} \chip; \chip))\;,
\end{equation}
where we abbreviated $S_{ab}$ as the sum of the first and second order terms:
\begin{equation}
 S_{ab}^{(1)} =  \Phi_{,ab} (\boldsymbol{n} \chi ; \chi), \quad
 S_{ab}^{R(2)} = -\frac{\chip}{\chi(\chi-\chip)}\Phi_{,ab} (\boldsymbol{n} \chip ; \chip) \frac{c}{H(a(\chi))} \Delta z (\chi)\;.
\label{eq::rsd}
\end{equation}
Clearly the first term recovers the usual lensing signal, while the second term accounts for a shift in redshift due to the following effects: Firstly the redshift changes due to the Sachs-Wolfe and the integrated Sachs-Wolfe effect \citep{Sachs1967}. Secondly it is altered by the peculiar motion of the source galaxies in the ambient LSS \citep{Kaiser1987,Hamilton1998},
\begin{equation}
	\Delta z = \Delta z_{\text{SW}} + \Delta z_{\text{ISW}} + \Delta z_{\text{V}}\;.
\end{equation}
The Sachs-Wolfe effect is the change on the emitted photon's redshift due to the gravitational potential at the source galaxy,
\begin{equation}\label{eq:SW}
	\Delta z_{\text{SW}}( \chi ) = \frac{ \Phi ( \boldsymbol{n} \chi ; \chi )}{ c^2}\;.
\end{equation}
The integrated Sachs-Wolfe effect describes the interaction of photons with an evolving gravitational potential along their line of sight. The line of sight fluctuation is given by \citep{Sachs1967}
\begin{equation}\label{eq:ISW}
	\Delta z_{\text{ISW}}( \chi ) = \frac{2}{c^3} \int_0^\chi \mathrm{d} \chi'  \frac{\partial}{\partial \eta} \Phi( \boldsymbol{n} \chi'; \chi')\;,
\end{equation}
which vanishes for matter dominated epochs, since $\Phi = \mathrm{const}$ in this case.
Finally, we also consider the peculiar motion contribution to the observed redshift \citep{Kaiser1987},
\begin{equation}
	\Delta z_{\text{V}}( \chi ) = \frac{\upsilon_\parallel( \chi )}{c}\;,
\end{equation}
where $\upsilon_\parallel$ is the peculiar velocity component of the source galaxy along the line of sight. Here, we compute only the peculiar motion contribution to the galaxy redshift, while effects on propagation of the light bundle due to a moving source are neglected \citep{Bonvin2008}. The three effects can partially cancel each other, but they can also add up, epecially between the Sachs-Wolfe effect and the peculiar motion a strong correlation exists in the sense that galaxies located in deep potential wells having a high velocity dispersion due to the virial theorem.

\subsection{Second-order corrections to the light propagation}
A common approximation made in gravitational lensing is that the gravitational potentials involved are small, i.e. $\Phi\ll c^2$. Lensing is thus studied in an effective Minkowskian spacetime due to the conformal invariance off null geodesics where the perturbations are linear in the Newtonian potential. Thus the effective speed of light is expanded up to first order. However, higher order terms affect the lensing signal \citep{PhysRevD.81.083002, giblin_jr_general_2017, tansella_full-sky_2017}:
\begin{equation}
 c'= c \sqrt{\frac{1-\frac{2\Phi}{c^2}}{1+\frac{2\Phi}{c^2}}}\approx c - \frac{2 \Phi}{c} + \frac{2\Phi^2}{c^3}\; .
\end{equation}
The second order term generates also second order correction to the lensing potential, $ S_{ab}^{P(2)}$ , that gets ultimately summed up with \eqref{eq::rsd},
\begin{equation}
\label{eq::velocity}
 S_{ab}^{P(2)} = -\frac{1}{c^2}\Phi^2_{,ab}\; .
\end{equation}

\subsection{Temporal Born-effect}
The Born-approximation sets the photon path to be an idealized straight FLRW-geodesic \citep{Schneider1988, Lee1990,2001PhR...340..291B}. Using the actual geodesics of the photons complicate matters significantly, since new positions must be computed from past deflected ones \citep{Petri2017}. This spatial aspect of the Born-approximation has been widely studied before analytically \citep{Cooray:2002mj, Shapiro:2006em, Krause:2009yr, schafer_validity_2012, petri_validity_2016} or in numerical simulations \citep{Hilbert2009, giblin_jr_general_2017}, and fond to be small.

Similarly, the case of lensing of the cosmic microwave background has been treated analytically \citet{hagstotz_born-corrections_2014, pratten_impact_2016, Marozzi2016} and through simulations \citep{carbone_lensed_2009, calabrese_multiple_2015}, with the particular implication for changing the distance to the last-scattering surface \citep{bonvin_we_2015,clarkson_what_2014}. Nonetheless, this is not its only consequence. Let us assume that photons move with the effective speed of the perturbed metric but follow radial geodesics of the unperturbed FRLW geometry. A photon following perturbed geodesics has a varying effective velocity, sometimes it overtakes an idealised photon that follows a FLRW geodesic and other times it gets left behind. Consequently, they would encounter the same structure at different evolutionary stages. Since this time difference depends on the potentials the photon underwent before, it is
also an integrated effect.

The growth of potentials is determined by the factor $D_+/a$, which we expand up to second order to quantify the temporal Born-correction,
\begin{equation}
 \left . \frac{D_+(a)}{a}  \right|  \approx \frac{D_+(a)}{a} + \frac{\mathrm{d}}{\mathrm{d}a} \left.\left(\frac{D_+(a)}{a}\right)\right|_a \Delta a\; ,
\end{equation}
where $\Delta a = a^2H(a)\Delta t$, and the time departure is given by the difference in effective
light speed from the lens to the source,
\begin{equation}
 \Delta t = \int_\chip^\chi \frac{\mathrm{d}\chi^{\prime\prime}}{c^{\prime}}-\int_\chip^\chi \frac{\mathrm{d}\chi^{\prime\prime}}{c} \approx 2 \int_\chip^\chi \mathrm{d}\chi^{\prime\prime}
\frac{\Phi(\boldsymbol{n} \chi^{\prime\prime} ;\chi^{\prime\prime})}{c^3}\; .
\end{equation}
Finally the temporal Born-approximation adds a second order contribution to \eqref{eq::velocity},
\begin{equation}
  \label{eq::born}
	S_{ab}^{B(2)} = \frac{2}{c^2} \frac{\mathrm{d}}{\mathrm{d} \eta}\left(\frac{D_+(a)}{a}\right)_\chi \Phi_{,ab}(\boldsymbol{n} \chi; 0) \int_\chip^\chi \mathrm{d}\chi''\Phi(\boldsymbol{n}\chi^{\prime\prime};\chi^{\prime\prime})\; ,
\end{equation}
which is also vanishing in the matter dominated epoch of the Universe similarly to the iSW effect, since both effects have a very similar origin, again due to the null property of photons.

\subsection{Gravitomagnetic effect}
The most general energy-momentum tensor compatible with the cosmological symmetries, is the energy-momentum tensor of a perfect fluid, for which one finds
\begin{equation}
	T^{\alpha \beta} = ( \rho c^2 + p ) \upsilon^{\alpha} \upsilon^{\beta} - p g^{\alpha \beta}\; ,
\end{equation}
where $ \rho $ is the mass density and $p$ the fluids pressure, both measured in a reference frame with a normalized 4-velocity $ \upsilon^{\alpha} = ( \upsilon^{0}, \boldsymbol{\upsilon} )$.
Under the assumption that the gravitational lenses are slowly moving, the kinetic contribution to gravity can be ignored and the dominant component is $T^{00} = \rho c^2 $. We study the validity of this assumption \citep{sereno_gravitational_2003, schafer_weak_2006}, considering the contribution of $ T^{0i} = c \rho \upsilon^{i}$. It can be shown that the effective speed of light gains an additional term \citep{Dodelson2005}, 
\begin{equation}
	c' = c - \frac{2 \Phi}{c} + \frac{4}{ c^{2}} A_{\parallel}\;,
\end{equation}
where the line of sight component of the vectorial mode is given by
\begin{equation}
A_{\parallel} ( \boldsymbol{x} , t ) = -G \int \frac{ \rho( \boldsymbol{x}\,', t) \upsilon_{\parallel} ( \boldsymbol{x}\,',t)}{| \boldsymbol{x} - \boldsymbol{x} \,'|} d \boldsymbol{x}' \;.
\end{equation}
As a consequence, the second order correction to the lensing potential is 
\begin{equation}
  \label{eq::gm_source}
	S_{ab}^{G(2)} = - \frac{2}{c} A_{\parallel}( \boldsymbol{n} \chi ; \chi )\;.
\end{equation}
\citet{Dodelson2005} investigated this effect analytically along with Born-corrections and lens-lens coupling, and it was found to be small.

\section{Corrections to weak lensing spectra}\label{sect_corrections}
We will now calculate the corrections to the weak lensing convergence spectrum (\ref{eq:Ckappa_ij}) subject to the effects described in \autoref{sect_corrections}. For this purpose, we write the lensing potential weighted by a probability distribution $p( z )$ in the $i$-th tomographic bin as 
\begin{equation}
	\overline{\psi}_i( \boldsymbol{n} ) = 
	\int_0^{\chi_H} \dd\mathrm{\chi}\: p_i(z) \frac{\dd z}{\dd \chi} \psi ( \boldsymbol{n} , \chi ) = 
	\frac{2}{c^2} \int_0^{\chi_H} \dd\mathrm{\chi} \int_{\chi}^{\chi_H} \dd \chi'\: g_i( \chi , \chi') S_{ab}( (\boldsymbol{n} \chi'; \chi'), (\boldsymbol{n} \chi ; \chi) )\;.
\end{equation}
Accordingly, the averaged convergence is
\begin{equation}
	\overline{\kappa}_i ( \boldsymbol{n} ) =  
	\frac{1}{c^2} \int_0^{\chi_H} \dd\mathrm{\chi} \int_{\chi}^{\chi_H} \dd \chi'\: g_i( \chi , \chi') S_{aa}[ (\boldsymbol{n} \chi'; \chi'), (\boldsymbol{n} \chi ; \chi) ]\;.
\end{equation}
The corresponding convergence power spectrum is
\begin{equation}
	\begin{split}
		C_{ij} (\ell) &= \int \frac{\dd^2 \ell'}{(2 \pi )^2} \left\langle \hat{\overline{\kappa}}_{i} ( \boldsymbol{\ell} ) \hat{\overline{\kappa}}^*_{j}( \boldsymbol{\ell} ') \right\rangle\\
					  &= \frac{1}{c^4} \int \frac{\dd^2 \ell'}{(2 \pi )^2} \int_0^{\chi_H} \dd \chi_1 \int_{\chi_1}^{\chi_H}\dd \chi_1' g_i( \chi_1, \chi_1') \int_0^{ \chi_H} \dd \chi_2 \int_{ \chi_2}^{ \chi_H} \dd \chi_2' g_j( \chi_2, \chi_2') \left\langle \hat{S}_{aa} ( \boldsymbol{\ell}, \chi_1,\chi_1') \hat{S}^*_{aa} ( \boldsymbol{\ell} ' , \chi_2, \chi_2')\right\rangle  \;,
	\end{split}
\end{equation}
where a hat denotes the two dimensional Fourier transform, we will not use this notation in the following if the argument to the Fourier variable is given and no confusion arises. In the computation of the convergence power spectra, two approximations have been made: The first corresponds to the flat sky approximation, valid for small angles, where we expect a stronger correlation between distorted images. The second one is the Limber approximation, in which we ignore any correlation along the line of sight, implying that the power spectrum of density fluctuations $P_\delta( k)$ can be evaluated at $k=\ell/ \chi$. For more information on common approximations made in cosmic shear we refer to \citet{Kitching2017}.

Furthermore we note that second order corrections involve products of two statistical fields in position space, which will give rise to convolutions in Fourier space. In the following we present the obtained autocorrelation power spectra for the different effects, and refer the reader to the appendix for the cross-correlations for sake of readability.

\subsection{Sachs-Wolfe effect}
The two dimensional power spectrum of the averaged convergence for the Sachs-Wolfe correction is, using Eq. (\ref{eq:SW}),
\begin{equation}
\label{eq::swcorrelation}
C_{ii}^{\mathrm{S}}( \ell ) = \frac{1}{c^6}\int_0^{\chi_H} \frac{\dd\chi }{\chi^2}\int \frac{\dd^2\ell^\prime}{(2\pi)^2} (\ell^\prime)^4 M_{ij}(\ell^\prime, |\vecdlp| ; \chi),
\end{equation}
where we abbreviated the mode coupling integral:
\begin{equation}
 M^{\mathrm{S}}_{ij}(\ell, \ell^\prime ; \chi) = \int_{\chi}^{\chi_H} \frac{\dd \chi^\prime}{(\chi^\prime)^2} d_i(\chi^\prime)d_j(\chi^\prime)P_{\Phi}\left(\frac{\ell}{\chi};\chi \right)P_{\Phi}\left(\frac{\ell^\prime}{\chi^\prime};\chi^\prime\right).
\end{equation}
with $d_i( \chi') = \frac{p_i( \chi')}{H(a( \chi'))} \frac{1}{ \chi'^2 }$. Clearly, we couple the potential power spectrum and two different scales and two different times. Due to the co-moving distances in the denominator this effect will be important only on large scales.

\subsection{Peculiar velocities}
In the case of peculiar motions only the projection of the velocity along the line of sight introduces a correction. In this case the Limber approximation can not be applied directly, since it assumes that there is no correlation between parallel modes. The final result is,
\begin{equation}
C^{\mathrm{V}}_{ij}(\ell) = \frac{1}{c^{4}}\int_0^{\chi_H} \frac{\dd\chi }{\chi^2} \int\frac{\dd k^\prime}{2\pi^2} M_{ij}(k^\prime, \ell; \chi).
\end{equation}
with mode coupling integral,
\begin{equation}
M^{\mathrm{V}}_{ij}(k^\prime, \ell; \chi) =
\chi^2\int_{\chi}^{\chi_H} \dd \chi^\prime\:\frac{\mathrm{d} D_+(a)}{\mathrm{d}\eta}  d_i(\chi^\prime) 
\int_\chi^{\chi_H}\dd \chi^{\prime\prime}\: \frac{\mathrm{d} D_+(a)}{\mathrm{d}\eta}  d_j(\chi^{\prime\prime})
P_{\delta}\left(k^\prime;\chi \right)
P_{\delta}\left(\frac{\sqrt{\ell^2+(k^\prime\chi^\prime)^2)}}{\chi};0\right)
j_0^{\prime \prime}(\kp|\chi^\prime-\chi^{\prime\prime}|)
\end{equation}
Comparing this correction to the one produced by the Sachs-Wolfe effect \eqref{eq::swcorrelation}, we find two structural differences even though both effects are local. Firstly, the convolution is restricted to modes 
perpendicular to $\boldsymbol{\ell}$. Secondly, there is no delta function between the primed comoving distances, instead we find a Bessel function which is a broader kernel, since velocities are originated by gradients of
potentials and therefore their correlation length is larger.

\subsection{Integrated Sachs-Wolfe effect}
For the integrated Sachs-Wolfe we obtain two different terms,
\begin{equation}
\label{eq::iswcorrelation}
C^{\mathrm{I}}_{ij}(\ell) = \frac{4}{c^{8}}\int_0^{\chi_H} \frac{\dd\chi }{\chi^2}\int \frac{\dd^2\ell^\prime}{(2\pi)^2} \left((\ell^\prime)^4 M_{1,\, ij}(\ell^\prime, |\vecdlp| ; \chi)+ (\ell^\prime)^2|\vecdlp|^2 M_{2,\, ij}(\ell^\prime, |\vecdlp| ; \chi)\right).
\end{equation}
with two mode coupling integrals
\begin{equation}
M^{\mathrm{I}}_{1,\, ij}(\ell, \ell^\prime ; \chi) =
\int_{\chi}^{\chi_H} \dd \chi_1^\prime d_i(\chi_1^\prime)
\int_\chi^{\chi_H}\dd \chi_2^\prime d_j(\chi_2^\prime)
\int_{\chi}^{\min(\chi_1^\prime,\chi_2^\prime)}
\frac{ \dd\chi ^{\prime \prime}}{ (\chi^{\prime\prime})^2}\left(\frac{\mathrm{d}}{\mathrm {d}\eta}\left(\frac{D_+(a)}{a}\right)\right)^2_{\chi^{\prime\prime}}P_{\Phi}\left(\frac{\ell}{\chi}; \chi \right)P_{\Phi}\left(\frac{\ell^\prime}{\chi^{\prime\prime}};0\right)
\end{equation}
and
\begin{equation}
\begin{split}
M^{\mathrm{I}}_{2,\, ij}(\ell, \ell^\prime ; \chi)&  =
\left(\frac{D_+(a)}{a}\right)_\chi \frac{\mathrm{d}}{\mathrm{d} \eta} \left( \frac{D_+(a)}{a} \right)_\chi
\int_{\chi}^{\chi_H} \dd \chi_1^\prime d_i(\chi_1^\prime)\int_\chi^{\chi_H}\dd \chi_2^\prime d_j(\chi_2^\prime)
\\ &\times 
\int_{\chi}^{\min(\chi_1^\prime,\chi_2^\prime)}
\frac{ \dd\chi ^{\prime \prime}}{ (\chi^{\prime\prime})^2}\left(\frac{D_+(a)}{a}\right)_{\chi^{\prime\prime}} \frac{\dd}{\dd \eta} \left( \frac{D_+(a)}{a} \right)_{ \chi''}P_{\Phi}\left(\frac{\ell}{\chi}; 0 \right)P_{\Phi}\left(\frac{\ell^\prime}{\chi^{\prime\prime}};0\right)\; .
\end{split}
\end{equation}
Both integrals depend on the change of the growth factor and thus are sensitive to the time evolution of the potentials opposed to Eq. (\ref{eq::swcorrelation}), due to the integration of the effect along the line of sight. However the scaling compared to Eq. (\ref{eq::swcorrelation}) is similar so that we expect the effect to be highest on very large scales.

\subsection{Second order corrections to light propagation}
Since the second order correction to the light-propagation only involves additional auto-correlations of the potential, i.e. correlations at the same positions or comoving distance, the Limber approximation can be applied directly after applying Wicks theorem to compute the correlation in the limit of Gaussian fields. Therefore we get:
\begin{equation}
C^{\mathrm{P}}_{ij}(\ell) = \frac{2 \ell ^4}{c^{8}}\int_0^{\chi_H} \frac{\dd\chi }{\chi^2}   \left( \int \frac{\dd^3 k^\prime}{(2\pi)^3} M_{ij}(k^\prime, |\vecdkp|; \chi)\right)_{k=\frac{\ell}{\chi}}\; ,
\end{equation}
with mode coupling integral,
\begin{equation}
 M^{\mathrm{P}}_{ij}(k, k^\prime ; \chi) = \int_{\chi}^{\chi_H} \dd \chi^\prime g_i(\chi,\chi^\prime)\int_\chi^{\chi_H}\dd \chi^{\prime\prime} g_j(\chi,\chi^{\prime\prime}) P_{\Phi}\left(k;\chi \right)P_{\Phi}\left(k^\prime;\chi\right)
\end{equation}
It should be noted that the mode coupling integral depends on the wave-vector directly and that $k = \ell/\chi$ is only inserted afterwards as opposed to the other corrections. The reason for this is exactly that we only apply the Limber projection in the very last step, as for normal cosmic shear power spectra.

\subsection{Temporal Born-approximation}
The temporal Born-corrections works very similar to the iSW corrections. We find: 
\begin{equation}
C^{\mathrm{B}}_{ij}(\ell) = \frac{4}{c^{10}}\int_0^{\chi_H} \frac{\dd\chi }{\chi^2}\left(\frac{\mathrm{d}}{\mathrm{d} \eta} \left( \frac{D_+(a)}{a} \right)\right)_\chi^2\int \frac{\dd^2\ell^\prime}{(2\pi)^2} (\ell^\prime)^4 M_{ij}(\ell^\prime, |\vecdlp| ; \chi)\; ,
\end{equation}
with mode coupling integral,
\begin{equation}
M^{\mathrm{B}}_{ij}(\ell, \ell^\prime ; \chi) =
\int_{\chi}^{\chi_H} \dd \chi_1^\prime g_i(\chi,\chi_1^\prime)\int_\chi^{\chi_H}\dd \chi_2^\prime g_j(\chi,\chi_2^\prime) \int_{\chi} ^{\min(\chi_1^\prime,\chi_2^\prime)} \frac{ \dd\chi ^{\prime \prime}}{ (\chi^{\prime\prime})^2}P_{\Phi}\left(\frac{\ell}{\chi};0 \right)P_{\Phi}\left(\frac{\ell^\prime}{\chi^{\prime\prime}};\chi^{\prime\prime}\right)\;.
\end{equation}
Compared to the other effects we already see that it will be weaker due to the prefactor of $c^{-10}$ and it will be highest on large scales as well like the SW and iSW contribution.

\subsection{Gravitomagnetic corrections}
Finally we investigate the effect of moving lenses and thus the autocorrelation of the gravitomagnetic correction:
\begin{equation}
C_{ij}^{\mathrm{G}}( \ell ) = \frac{4}{c^6}\int_0^{\chi_H} \dd\chi \left(\chi \frac{\mathrm{d}D_+(a)}{\mathrm{d}\eta}\right)^2\left[\int \frac{\dd k^\prime}{(2\pi)^2}|\vecdkp|^4 M_{ij}(k^\prime, |\vecdkp| ; \chi)\right]_{k=l/ \chi }\; ,
\end{equation}
where the mode coupling integral is given by
\begin{equation}
 M^{\mathrm{G}}_{ij}(k^\prime, |\vecdkp| ; \chi) = \int_{\chi}^{\chi_H} \dd \chi^\prime g_i(\chi,\chi^\prime)\int_{\chi}^{\chi_H} \dd \chi^{\prime\prime}g_j(\chi,\chi^{\prime\prime})P_{\delta}\left(k;0 \right)\int_{-1}^{1} \dd \mu\left( \mu^2 + \frac{k^\prime \mu (k -k^\prime \mu )}{|\vecdkp|^2  }\right) P_{\Phi}\left(|\vecdkp|;\chi \right)\; .
\end{equation}
Which will have the largest impact of the computed effects, which can be again seen by considering the pre-factor, which is only $c^{-6}$ here. Furthermore, it should be noted that the integrand of the mode coupling function depends of the orientation of the different wave vectors for which the correlation is formed, which is then averaged over. 

\subsection{Results}

\begin{figure}
\begin{center}
\includegraphics[width=0.5\textwidth]{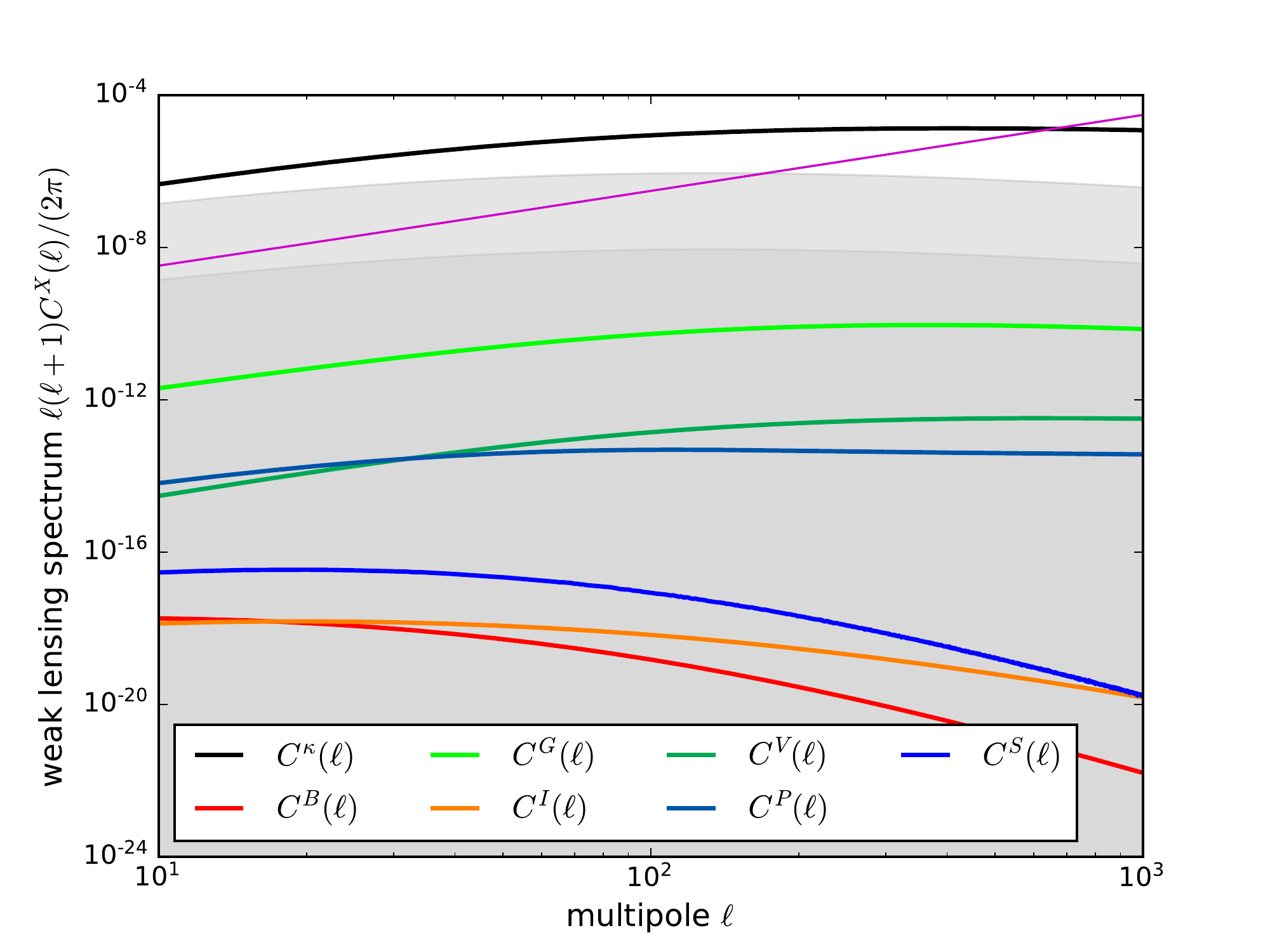}
\caption{Angular auto-spectra $C^{X}(\ell)$ for 2d weak lensing: temporal Born-effect (red) and the integrated Sachs-Wolfe effect (orange), and the result for linear weak lensing (black) in comparison, Euclid's shape noise (solid line for the actual value $\sigma_\epsilon^2/\bar{n}$, dashed line for $10^{-2}\sigma_\epsilon^2/\bar{n}$, magenta) and the cosmic variance limit $\Delta C^{\kappa}(\ell) = \sqrt{2/(2\ell+1)}C^{\kappa}(\ell)$ (grey bands for the actual cosmic variance and $10^{-2}$) of that value). In addition, the corrections due to gravitomagnetic effects ($C^G(\ell)$ in green), peculiar velocities ($C^V(\ell)$, dark green), higher-order corrections to the speed of propagation ($C^S(\ell)$, blue) are plotted.}
\label{fig:autocorrelations}
\end{center}
\end{figure}

\begin{figure}
\begin{center}
\includegraphics[width=1.0\textwidth]{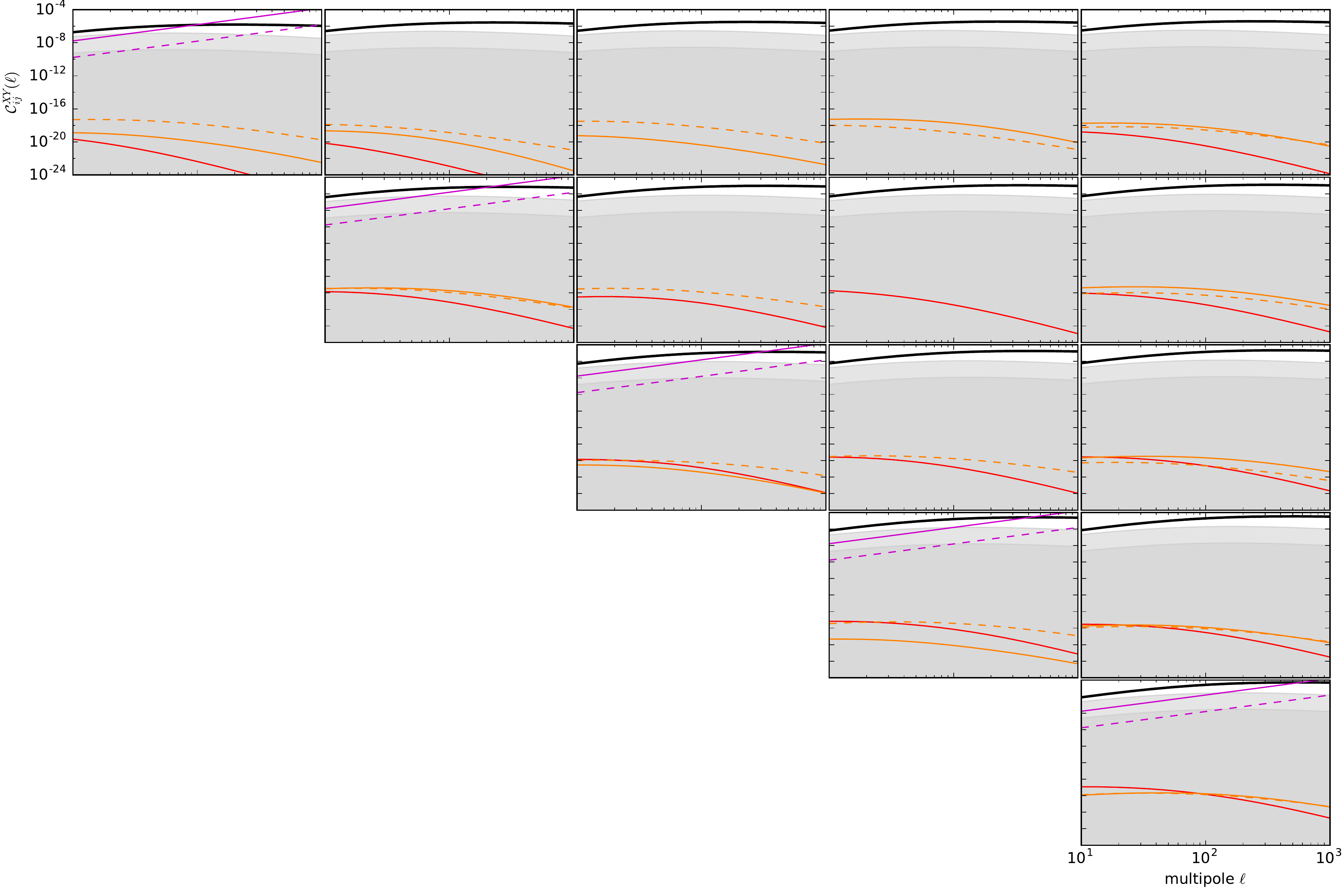}
\caption{Angular spectra $C^{XY}_{ij}(\ell)$ for 5-bin tomography in the representation $\mathcal{C}^{XY}_{ij}(\ell) = \ell(\ell+1)C^{XY}_{ij}(\ell)/(2\pi)$: temporal Born-effect (solid line, red) and the integrated Sachs-Wolfe effect (dashed line, orange) and the cross-correlation between the two effects. For comparison we show the result for linear weak lensing (solid line, black), Euclid's shape noise (solid line for the actual value $\sigma_\epsilon^2n_\mathrm{bin}/\bar{n}$, dashed line for $10^{-2}\sigma_\epsilon^2n_\mathrm{bin}/\bar{n}$, magenta) and the cosmic variance limit $\Delta C^{\kappa}_{ij}(\ell) = \sqrt{2/(2\ell+1)}C^{\kappa}_{ij}(\ell)$ (grey bands for the actual cosmic variance and $10^{-2}$ of that value).}
\label{fig:crosscorrelations1}
\end{center}
\end{figure}

\begin{figure}
\begin{center}
\includegraphics[width=1.0\textwidth]{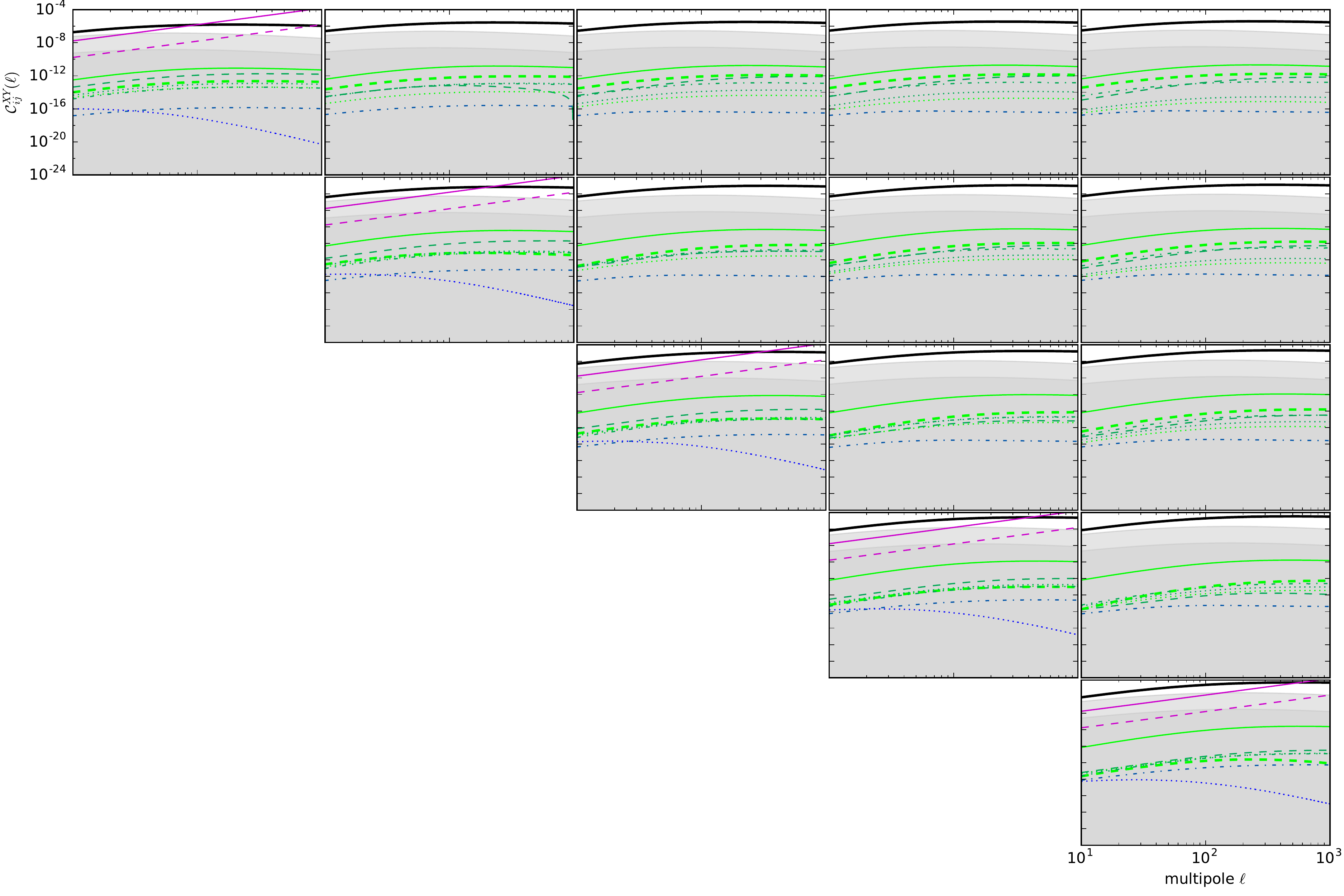}
\caption{Angular spectra $C^{XY}_{ij}(\ell)$ for 5-bin tomography in the representation $\mathcal{C}^{XY}_{ij}(\ell) = \ell(\ell+1)C^{XY}_{ij}(\ell)/(2\pi)$: gravitomagnetic corrections (solid line, green) and peculiar velocity corrections (dashed line, light green), corrections due to post-Newtonian potentials (dash-dotted line, light blue), Sachs-Wolfe effect (dotted line, blue) and all cross-correlations. For comparison we show the result for linear weak lensing (solid line, black), Euclid's shape noise (solid line for the actual value $\sigma_\epsilon^2n_\mathrm{bin}/\bar{n}$, dashed line for $10^{-2}\sigma_\epsilon^2n_\mathrm{bin}/\bar{n}$, magenta) and the cosmic variance limit $\Delta C^{\kappa}_{ij}(\ell) = \sqrt{2/(2\ell+1)}C^{\kappa}_{ij}(\ell)$ (grey bands for the actual cosmic variance and $10^{-2}$ of that value).}
\label{fig:crosscorrelations2}
\end{center}
\end{figure}

In \autoref{fig:autocorrelations} the auto-correlations calculated in the last section are shown. The black line shows the linear angular power spectrum while the magenta line represents the shape-noise contribution, the grey band represents cosmic variance. Clearly the corrections lie well below the shape noise and within the effect of cosmic variance. The largest effect stamps from gravitomagnetic effects. \autoref{fig:crosscorrelations1} and \autoref{fig:crosscorrelations2} show the cross-correlations (see \autoref{app:1}) between the different effects again shown together with the lensing spectrum, cosmic variance and shape noise. 

We calculate the Fisher matrix
\begin{equation}
F_{\mu\nu} = f_\mathrm{sky}\sum_\ell\frac{2\ell+1}{2}
\trace\left(\boldsymbol{C}^{-1}(\ell)\partial_\mu \boldsymbol{C}(\ell)\:
\boldsymbol{C}^{-1}(\ell)\partial_\nu \boldsymbol{C}(\ell)\right),
\end{equation}
where $\boldsymbol{C}$ is the tomographic covariance matrix consisting of the sum lensing power spectrum the shape noise and all corrections terms. The sky coverage for \textit{Euclid} is roughly $15000\;\mathrm{deg}^2$.
Next we are trying to fit a true model, $\boldsymbol{C}$, with a wrong model $\boldsymbol{C}_f$, not including the corrections. The bias $\delta_\mu$ is given by
\begin{equation}
\delta_\mu = (\boldsymbol{G}^{-1})_{\mu\nu} a^\nu\;,
\end{equation}
where 
\begin{equation}
a^\nu = \left\langle\frac{\partial L_f}{\partial \theta_\nu}\right\rangle , \quad G_{\mu\nu} = -\left\langle\frac{\partial^2 L_f}{\partial \theta^\mu\partial\theta^\nu}\right\rangle\; ,
\end{equation}
to be evaluated at the true model. $L_f$ refers to the log-likelihood of the false model. The parameter estimation process for a $w$CDM-model from a tomographic survey with 5 bins and the anticipated redshift distribution of Euclid's weak lensing data, we arrive at typical systematic errors of $\simeq 10^{-5}$ for $\Omega_m$, $\sigma_8$, $h$ and $w$, and of the order $\simeq10^{-6}$ for $n_s$, which is certainly well below the statistical error, consistent with the absolute values found for the correction.

\section{Summary}\label{sect_summary}
The subject of our investigation were gravitational secondary contributions to the weak lensing signal. These include as local effects $(i)$ the Sachs-Wolfe effect due to the non-zero gravitational potential where the lensed galaxy is situated, $(ii)$ contributions to the total redshift if the lensed galaxy has a non-zero peculiar velocity relative to the Hubble-flow, corrections due to general relativity in the weak field limit because of $(iii)$ quadratic corrections to the effective speed of propagation and $(iv)$ gravitomagnetic terms due to the contributions of the momentum density to the gravitational field. As integrated effects, we considered $(v)$ the integrated Sachs-Wolfe effect affecting the redshift of the lensed galaxy and, in addition, we evaluate the effect of the non-uniform effective speed of light in gravitational potentials that gives rise to an equivalent correction corresponding to Born-corrections. As photons are travelling along null-geodesics from a source galaxy to the observer, their effective speed of propagation is modulated by the depth of the gravitational potentials that they need to traverse. Consequently, they encounter deflecting structures at a different time and therefore at a different stage of structure formation in comparison to idealised photons which follow null-geodesics of a FLRW-spacetime. 

These corrections are computed for a FLRW-cosmology with weak perturbations which source gravitational potentials that are effectively Newtonian. Structure formation was treated in the linear limit, which enforces near-Gaussian statistics of the gravitational potential fluctuations. From this model of linear and Gaussian perturbations we derive angular spectra of all correction terms and their cross-correlations in perturbation theory, where the nonlinear dependences of all effects on the fundamental fields would cause non-Gaussian statistical properties. We carry out our computation for the characteristics of the Euclid-survey, but a similar strength of the correction terms for the weak lensing signal should be applicable for any reasonably deep weak lensing survey. 

There is the general tendency that effects related to peculiar velocities, either of the source galaxies leading to a change in redshift relative to the cosmological one or of the lensing structure giving rise to gravitomagnetic effects, provide the largest corrections, followed by effects involving the gravitational potential at the lens, i.e. higher-order corrections to the light propagation, or due to the large-scale structure into which the source galaxy is embedded, causing a gravitational redshift. The smallest effects are integrated effects depending on the evolution of the gravitational potentials, which are interpreted as an integrated Sachs-Wolfe effect or the temporal Born-correction. The latter two effects show the same phenomenology in terms of cosmological parameters and would be absent in a flat, matter-dominated cosmology with $\Omega_\mathrm{m}=1$. There, the growth function $D_+(a)$ is equal to the scale factor $a$ and consequently, fluctuations in the gravitational potential are constant and do not give rise to integrated effects. We find the conceptual difference between these two effects striking, in particular because they give at the same time rise to very similar expressions and show identical dependences on cosmological parameters.

The computation was done for a $w$CDM-cosmology and the magnitude of the correction terms in comparison to the linear weak lensing is $10^{-5\ldots-6}$ at most, indicating that the most important secondary effects in weak lensing are in fact reduced shear and magnification corrections \citep{Krause:2009yr} and intrinsic alignments \citep{blazek_tidal_2015, joachimi_galaxy_2015, schaefer_angular_2015, kiessling_galaxy_2015, kirk_galaxy_2015, troxel_intrinsic_2015}. We would argue that the magnitude of the effects does not strongly depend on the particular dark energy model and should be valid for a $\Lambda$CDM-cosmology as well. Due to the smallness of the individual effects in comparison to the linear weak lensing signal and the shape noise implies that there should be a negligible effect on the estimation of cosmological parameters. We compute the resulting systematic errors and found them to be very small.

\section*{Acknowledgements}
We would like to thank P. Norberg for giving us the opportunity to write up this project. RR acknowledges funding by the graduate college \textit{Astrophysics of cosmological probes of gravity} by Landesgraduiertenakademie Baden-W\"urttemberg.

\bibliographystyle{mnras}
\bibliography{references}

\begin{thebibliography}{}
\makeatletter
\relax
\def\mn@urlcharsother{\let\do\@makeother \do\$\do\&\do\#\do\^\do\_\do\%\do\~}
\def\mn@doi{\begingroup\mn@urlcharsother \@ifnextchar [ {\mn@doi@}
  {\mn@doi@[]}}
\def\mn@doi@[#1]#2{\def\@tempa{#1}\ifx\@tempa\@empty \href
  {http://dx.doi.org/#2} {doi:#2}\else \href {http://dx.doi.org/#2} {#1}\fi
  \endgroup}
\def\mn@eprint#1#2{\mn@eprint@#1:#2::\@nil}
\def\mn@eprint@arXiv#1{\href {http://arxiv.org/abs/#1} {{\tt arXiv:#1}}}
\def\mn@eprint@dblp#1{\href {http://dblp.uni-trier.de/rec/bibtex/#1.xml}
  {dblp:#1}}
\def\mn@eprint@#1:#2:#3:#4\@nil{\def\@tempa {#1}\def\@tempb {#2}\def\@tempc
  {#3}\ifx \@tempc \@empty \let \@tempc \@tempb \let \@tempb \@tempa \fi \ifx
  \@tempb \@empty \def\@tempb {arXiv}\fi \@ifundefined
  {mn@eprint@\@tempb}{\@tempb:\@tempc}{\expandafter \expandafter \csname
  mn@eprint@\@tempb\endcsname \expandafter{\@tempc}}}

\bibitem[\protect\citeauthoryear{{Amara} \& {R{\'e}fr{\'e}gier}}{{Amara} \&
  {R{\'e}fr{\'e}gier}}{2007}]{2007MNRAS.381.1018A}
{Amara} A.,  {R{\'e}fr{\'e}gier} A.,  2007, \mn@doi [MNRAS]
  {10.1111/j.1365-2966.2007.12271.x}, \href
  {http://esoads.eso.org/abs/2007MNRAS.381.1018A} {381, 1018}

\bibitem[\protect\citeauthoryear{Amara \& R{\'e}fr{\'e}gier}{Amara \&
  R{\'e}fr{\'e}gier}{2008}]{amara_systematic_2008}
Amara A.,  R{\'e}fr{\'e}gier A.,  2008, \mn@doi [MNRAS]
  {10.1111/j.1365-2966.2008.13880.x}, 391, 228

\bibitem[\protect\citeauthoryear{{Bacon}, {Refregier}  \& {Ellis}}{{Bacon}
  et~al.}{2000}]{2000MNRAS.318..625B}
{Bacon} D.~J.,  {Refregier} A.~R.,   {Ellis} R.~S.,  2000, \mn@doi [MNRAS]
  {10.1046/j.1365-8711.2000.03851.x}, \href
  {http://adsabs.harvard.edu/abs/2000MNRAS.318..625B} {318, 625}

\bibitem[\protect\citeauthoryear{{Bartelmann}}{{Bartelmann}}{2010}]{2010CQGra..27w3001B}
{Bartelmann} M.,  2010, \mn@doi [Classical and Quantum Gravity]
  {10.1088/0264-9381/27/23/233001}, \href
  {http://esoads.eso.org/abs/2010CQGra..27w3001B} {27, 233001}

\bibitem[\protect\citeauthoryear{{Bartelmann} \& {Schneider}}{{Bartelmann} \&
  {Schneider}}{2001}]{2001PhR...340..291B}
{Bartelmann} M.,  {Schneider} P.,  2001, \physrep, \href
  {http://esoads.eso.org/cgi-bin/nph-bib_query?bibcode=2001PhR...340..291B&amp;db_key=PHY}
  {340, 291}

\bibitem[\protect\citeauthoryear{Bender et~al.,}{Bender et~al.}{2000}]{Bendera}
Bender R.,  et~al., 2000, in , Deep Fields.
Springer-Verlag, pp 96--101, \mn@doi{10.1007/10854354_18}

\bibitem[\protect\citeauthoryear{Bernardeau, Bonvin  \& Vernizzi}{Bernardeau
  et~al.}{2010}]{PhysRevD.81.083002}
Bernardeau F.,  Bonvin C.,   Vernizzi F.,  2010, \mn@doi [Phys. Rev. D]
  {10.1103/PhysRevD.81.083002}, 81, 083002

\bibitem[\protect\citeauthoryear{Blazek, Vlah  \& Seljak}{Blazek
  et~al.}{2015}]{blazek_tidal_2015}
Blazek J.,  Vlah Z.,   Seljak U.,  2015, JCAP, 08, 15

\bibitem[\protect\citeauthoryear{{Bolzonella}, {Miralles}  \&
  {Pell{\'o}}}{{Bolzonella} et~al.}{2000}]{Bolzonella2000}
{Bolzonella} M.,  {Miralles} J.-M.,   {Pell{\'o}} R.,  2000, \aap, \href
  {http://adsabs.harvard.edu/abs/2000A%26A...363..476B} {363, 476}

\bibitem[\protect\citeauthoryear{{Bonvin}}{{Bonvin}}{2008}]{Bonvin2008}
{Bonvin} C.,  2008, \mn@doi [\prd] {10.1103/PhysRevD.78.123530}, \href
  {http://adsabs.harvard.edu/abs/2008PhRvD..78l3530B} {78, 123530}

\bibitem[\protect\citeauthoryear{Bonvin, Clarkson, Durrer, Maartens  \&
  Umeh}{Bonvin et~al.}{2015}]{bonvin_we_2015}
Bonvin C.,  Clarkson C.,  Durrer R.,  Maartens R.,   Umeh O.,  2015, \mn@doi
  [JCAP] {10.1088/1475-7516/2015/06/050}, 2015, 050

\bibitem[\protect\citeauthoryear{{Bridle}}{{Bridle}}{2008}]{Briddle2008}
{Bridle} S.,  2008, in A Decade of Dark Energy.

\bibitem[\protect\citeauthoryear{Calabrese, Carbone, Fabbian, Baldi  \&
  Baccigalupi}{Calabrese et~al.}{2015}]{calabrese_multiple_2015}
Calabrese M.,  Carbone C.,  Fabbian G.,  Baldi M.,   Baccigalupi C.,  2015,
  \mn@doi [JCAP] {10.1088/1475-7516/2015/03/049}, 2015, 049

\bibitem[\protect\citeauthoryear{Carbone, Baccigalupi, Bartelmann, Matarrese
  \& Springel}{Carbone et~al.}{2009}]{carbone_lensed_2009}
Carbone C.,  Baccigalupi C.,  Bartelmann M.,  Matarrese S.,   Springel V.,
  2009, \mn@doi [MNRAS] {10.1111/j.1365-2966.2009.14746.x}, 396, 668

\bibitem[\protect\citeauthoryear{{Castro}, {Heavens}  \& {Kitching}}{{Castro}
  et~al.}{2005}]{2005PhRvD..72b3516C}
{Castro} P.~G.,  {Heavens} A.~F.,   {Kitching} T.~D.,  2005, \mn@doi [Phys.
  Rev. D] {10.1103/PhysRevD.72.023516}, \href
  {http://esoads.eso.org/abs/2005PhRvD..72b3516C} {72, 023516}

\bibitem[\protect\citeauthoryear{Clarkson, Umeh, Maartens  \& Durrer}{Clarkson
  et~al.}{2014}]{clarkson_what_2014}
Clarkson C.,  Umeh O.,  Maartens R.,   Durrer R.,  2014, \mn@doi [JCAP]
  {10.1088/1475-7516/2014/11/036}, 2014, 036

\bibitem[\protect\citeauthoryear{{Cooray} \& {Hu}}{{Cooray} \&
  {Hu}}{2002a}]{2002ApJ...574...19C}
{Cooray} A.,  {Hu} W.,  2002a, \mn@doi [\apj] {10.1086/340892}, \href
  {http://esoads.eso.org/abs/2002ApJ...574...19C} {574, 19}

\bibitem[\protect\citeauthoryear{Cooray \& Hu}{Cooray \&
  Hu}{2002b}]{Cooray:2002mj}
Cooray A.,  Hu W.,  2002b, \mn@doi [Astrophys. J.] {10.1086/340892}, 574, 19

\bibitem[\protect\citeauthoryear{Croft \& Metzler}{Croft \&
  Metzler}{2000}]{croft_weak-lensing_2000}
Croft R.~A.,  Metzler C.~A.,  2000, ApJ, 545, 561

\bibitem[\protect\citeauthoryear{Croft, Freeman, Schuster  \& Schafer}{Croft
  et~al.}{2017}]{Croft2017}
Croft R. A.~C.,  Freeman P.~E.,  Schuster T.~S.,   Schafer C.~M.,  2017,
  \mn@doi [MNRAS] {10.1093/mnras/stx1206}, 469, 4422

\bibitem[\protect\citeauthoryear{Cunha, Huterer, Lin, Busha  \& Wechsler}{Cunha
  et~al.}{2014}]{Cunha2014}
Cunha C.~E.,  Huterer D.,  Lin H.,  Busha M.~T.,   Wechsler R.~H.,  2014,
  \mn@doi [MNRAS] {10.1093/mnras/stu1424}, 444, 129

\bibitem[\protect\citeauthoryear{Dodelson, Kolb, Matarrese, Riotto  \&
  Zhang}{Dodelson et~al.}{2005}]{Dodelson2005}
Dodelson S.,  Kolb E.~W.,  Matarrese S.,  Riotto A.,   Zhang P.,  2005, \mn@doi
  [Physical Review D] {10.1103/physrevd.72.103004}, 72

\bibitem[\protect\citeauthoryear{Eisenstein \& Hu}{Eisenstein \&
  Hu}{1998}]{Eisenstein1998}
Eisenstein D.~J.,  Hu W.,  1998, \mn@doi [The Astrophysical Journal]
  {10.1086/305424}, 496, 605

\bibitem[\protect\citeauthoryear{Eisenstein \& Hu}{Eisenstein \&
  Hu}{1999}]{eisenstein_power_1999}
Eisenstein D.~J.,  Hu W.,  1999, \mn@doi [ApJ] {10.1086/306640}, 511, 5

\bibitem[\protect\citeauthoryear{Gazta{\~{n}}aga, Eriksen, Crocce, Castander,
  Fosalba, Marti, Miquel  \& Cabr{\'{e}}}{Gazta{\~{n}}aga
  et~al.}{2012}]{Gaztanaga2012}
Gazta{\~{n}}aga E.,  Eriksen M.,  Crocce M.,  Castander F.~J.,  Fosalba P.,
  Marti P.,  Miquel R.,   Cabr{\'{e}} A.,  2012, \mn@doi [MNRAS]
  {10.1111/j.1365-2966.2012.20613.x}, 422, 2904

\bibitem[\protect\citeauthoryear{Giblin~Jr, Mertens, Starkman  \&
  Zentner}{Giblin~Jr et~al.}{2017}]{giblin_jr_general_2017}
Giblin~Jr J.~T.,  Mertens J.~B.,  Starkman G.~D.,   Zentner A.~R.,  2017, ArXiv
  e-prints 1707.06640

\bibitem[\protect\citeauthoryear{Hagstotz, Schaefer  \& Merkel}{Hagstotz
  et~al.}{2015}]{hagstotz_born-corrections_2014}
Hagstotz S.,  Schaefer B.~M.,   Merkel P.~M.,  2015, MNRAS, 454, 831

\bibitem[\protect\citeauthoryear{Hamilton}{Hamilton}{1998}]{Hamilton1998}
Hamilton A. J.~S.,  1998, in , Astrophysics and Space Science Library.
Springer Netherlands, pp 185--275, \mn@doi{10.1007/978-94-011-4960-0_17}

\bibitem[\protect\citeauthoryear{{Heavens}}{{Heavens}}{2003}]{2003MNRAS.343.1327H}
{Heavens} A.,  2003, \mn@doi [MNRAS] {10.1046/j.1365-8711.2003.06780.x}, \href
  {http://adsabs.harvard.edu/abs/2003MNRAS.343.1327H} {343, 1327}

\bibitem[\protect\citeauthoryear{Hilbert, Hartlap, White  \& Schneider}{Hilbert
  et~al.}{2009}]{Hilbert2009}
Hilbert S.,  Hartlap J.,  White S. D.~M.,   Schneider P.,  2009, \mn@doi
  [Astronomy {\&} Astrophysics] {10.1051/0004-6361/200811054}, 499, 31

\bibitem[\protect\citeauthoryear{Hoekstra \& Jain}{Hoekstra \&
  Jain}{2008}]{hoekstra_weak_2008}
Hoekstra H.,  Jain B.,  2008, \mn@doi [Annual Review of Nuclear and Particle
  Science] {10.1146/annurev.nucl.58.110707.171151}, 58, 99

\bibitem[\protect\citeauthoryear{Hoekstra, Viola  \& Herbonnet}{Hoekstra
  et~al.}{2017}]{Hoekstra2017}
Hoekstra H.,  Viola M.,   Herbonnet R.,  2017, \mn@doi [MNRAS]
  {10.1093/mnras/stx724}, 468, 3295

\bibitem[\protect\citeauthoryear{{Hu}}{{Hu}}{1999}]{1999ApJ...522L..21H}
{Hu} W.,  1999, \mn@doi [ApJL] {10.1086/312210}, \href
  {http://esoads.eso.org/abs/1999ApJ...522L..21H} {522, L21}

\bibitem[\protect\citeauthoryear{Huterer}{Huterer}{2010}]{huterer_weak_2010}
Huterer D.,  2010, \mn@doi [Gen. Relat. Grav.] {10.1007/s10714-010-1051-z}, 42,
  2177

\bibitem[\protect\citeauthoryear{{Jain}, {Seljak}  \& {White}}{{Jain}
  et~al.}{2000}]{2000ApJ...530..547J}
{Jain} B.,  {Seljak} U.,   {White} S.,  2000, \mn@doi [\apj] {10.1086/308384},
  \href {http://esoads.eso.org/abs/2000ApJ...530..547J} {530, 547}

\bibitem[\protect\citeauthoryear{Joachimi et~al.,}{Joachimi
  et~al.}{2015}]{joachimi_galaxy_2015}
Joachimi B.,  et~al., 2015, \mn@doi [Space Science Reviews]
  {10.1007/s11214-015-0177-4}, 193, 1

\bibitem[\protect\citeauthoryear{Kaiser}{Kaiser}{1987}]{Kaiser1987}
Kaiser N.,  1987, \mn@doi [MNRAS] {10.1093/mnras/227.1.1}, 227, 1

\bibitem[\protect\citeauthoryear{{Kaiser}, {Wilson}  \& {Luppino}}{{Kaiser}
  et~al.}{2000}]{2000astro.ph..3338K}
{Kaiser} N.,  {Wilson} G.,   {Luppino} G.~A.,  2000, ArXiv e-prints 000338,
  \href {http://adsabs.harvard.edu/abs/2000astro.ph..3338K} {}

\bibitem[\protect\citeauthoryear{Kayo, Takada  \& Jain}{Kayo
  et~al.}{2013}]{kayo_information_2013}
Kayo I.,  Takada M.,   Jain B.,  2013, \mn@doi [MNRAS] {10.1093/mnras/sts340},
  429, 344

\bibitem[\protect\citeauthoryear{Kiessling et~al.,}{Kiessling
  et~al.}{2015}]{kiessling_galaxy_2015}
Kiessling A.,  et~al., 2015, \mn@doi [Space Science Reviews]
  {10.1007/s11214-015-0203-6}, 193, 67

\bibitem[\protect\citeauthoryear{Kilbinger}{Kilbinger}{2015}]{Kilbinger2015}
Kilbinger M.,  2015, \mn@doi [Reports on Progress in Physics]
  {10.1088/0034-4885/78/8/086901}, 78, 086901

\bibitem[\protect\citeauthoryear{Kilbinger et~al.,}{Kilbinger
  et~al.}{2009}]{kilbinger_dark-energy_2009}
Kilbinger M.,  et~al., 2009, \mn@doi [A \& A] {10.1051/0004-6361/200811247},
  497, 677

\bibitem[\protect\citeauthoryear{Kirk et~al.,}{Kirk
  et~al.}{2015}]{kirk_galaxy_2015}
Kirk D.,  et~al., 2015, \mn@doi [Space Science Reviews]
  {10.1007/s11214-015-0213-4}, 193, 139

\bibitem[\protect\citeauthoryear{Kitching et~al.,}{Kitching
  et~al.}{2014}]{kitching_3d_2014}
Kitching T.~D.,  et~al., 2014, MNRAS, 442, 1326

\bibitem[\protect\citeauthoryear{Kitching, Alsing, Heavens, Jimenez, McEwen  \&
  Verde}{Kitching et~al.}{2017}]{Kitching2017}
Kitching T.~D.,  Alsing J.,  Heavens A.~F.,  Jimenez R.,  McEwen J.~D.,   Verde
  L.,  2017, \mn@doi [MNRAS] {10.1093/mnras/stx1039}, 469, 2737

\bibitem[\protect\citeauthoryear{Krause \& Hirata}{Krause \&
  Hirata}{2010}]{Krause:2009yr}
Krause E.,  Hirata C.~M.,  2010, \mn@doi [Astron. Astrophys.]
  {10.1051/0004-6361/200913524}, 523, A28

\bibitem[\protect\citeauthoryear{{LSST Science Collaboration} et~al.,}{{LSST
  Science Collaboration} et~al.}{2009}]{Collaboration2009}
{LSST Science Collaboration} et~al., 2009, arXiv preprint arXiv:0912.0201L

\bibitem[\protect\citeauthoryear{{Laureijs} et~al.,}{{Laureijs}
  et~al.}{2011}]{EuclidStudyReport}
{Laureijs} R.,  et~al., 2011, preprint, \href
  {http://adsabs.harvard.edu/abs/2011arXiv1110.3193L} {} (\mn@eprint {arXiv}
  {1110.3193})

\bibitem[\protect\citeauthoryear{Lee}{Lee}{2011}]{lee_intrinsic_2011}
Lee J.,  2011, \mn@doi [ApJ] {10.1088/0004-637X/732/2/99}, 732, 99

\bibitem[\protect\citeauthoryear{{Lee} \& {Paczynski}}{{Lee} \&
  {Paczynski}}{1990}]{Lee1990}
{Lee} M.~H.,  {Paczynski} B.,  1990, \mn@doi [\apj] {10.1086/168888}, \href
  {http://adsabs.harvard.edu/abs/1990ApJ...357...32L} {357, 32}

\bibitem[\protect\citeauthoryear{{Limber}}{{Limber}}{1954}]{1954ApJ...119..655L}
{Limber} D.~N.,  1954, \apj, \href
  {http://esoads.eso.org/abs/1954ApJ...119..655L} {119, 655}

\bibitem[\protect\citeauthoryear{{Linder} \& {Jenkins}}{{Linder} \&
  {Jenkins}}{2003}]{2003MNRAS.346..573L}
{Linder} E.~V.,  {Jenkins} A.,  2003, \mn@doi [\mnras]
  {10.1046/j.1365-2966.2003.07112.x}, \href
  {http://esoads.eso.org/abs/2003MNRAS.346..573L} {346, 573}

\bibitem[\protect\citeauthoryear{{Marozzi}, {Fanizza}, {Di Dio}  \&
  {Durrer}}{{Marozzi} et~al.}{2016}]{Marozzi2016}
{Marozzi} G.,  {Fanizza} G.,  {Di Dio} E.,   {Durrer} R.,  2016, \mn@doi [JCAP]
  {10.1088/1475-7516/2016/09/028}, \href
  {http://adsabs.harvard.edu/abs/2016JCAP...09..028M} {9, 028}

\bibitem[\protect\citeauthoryear{Miller et~al.,}{Miller
  et~al.}{2013}]{Miller2013}
Miller L.,  et~al., 2013, \mn@doi [MNRAS] {10.1093/mnras/sts454}, 429, 2858

\bibitem[\protect\citeauthoryear{Percival, Samushia, Ross, Shapiro  \&
  Raccanelli}{Percival et~al.}{2011}]{Percival2011}
Percival W.~J.,  Samushia L.,  Ross A.~J.,  Shapiro C.,   Raccanelli A.,  2011,
  \mn@doi [Philosophical Transactions of the Royal Society A: Mathematical,
  Physical and Engineering Sciences] {10.1098/rsta.2011.0370}, 369, 5058

\bibitem[\protect\citeauthoryear{Petri, Haiman  \& May}{Petri
  et~al.}{2017a}]{petri_validity_2016}
Petri A.,  Haiman Z.,   May M.,  2017a, PRD, 95, 123503

\bibitem[\protect\citeauthoryear{{Petri}, {Haiman}  \& {May}}{{Petri}
  et~al.}{2017b}]{Petri2017}
{Petri} A.,  {Haiman} Z.,   {May} M.,  2017b, \mn@doi [\prd]
  {10.1103/PhysRevD.95.123503}, \href
  {http://adsabs.harvard.edu/abs/2017PhRvD..95l3503P} {95, 123503}

\bibitem[\protect\citeauthoryear{Pratten \& Lewis}{Pratten \&
  Lewis}{2016}]{pratten_impact_2016}
Pratten G.,  Lewis A.,  2016, \mn@doi [JCAP] {10.1088/1475-7516/2016/08/047},
  2016, 047

\bibitem[\protect\citeauthoryear{Sachs \& Wolfe}{Sachs \&
  Wolfe}{1967}]{Sachs1967}
Sachs R.~K.,  Wolfe A.~M.,  1967, \mn@doi [The Astrophysical Journal]
  {10.1086/148982}, 147, 73

\bibitem[\protect\citeauthoryear{Schaefer \& Merkel}{Schaefer \&
  Merkel}{2015}]{schaefer_angular_2015}
Schaefer B.~M.,  Merkel P.~M.,  2015, ArXiv e-prints 1506.07366

\bibitem[\protect\citeauthoryear{Sch{\"a}fer \& Bartelmann}{Sch{\"a}fer \&
  Bartelmann}{2006}]{schafer_weak_2006}
Sch{\"a}fer B.~M.,  Bartelmann M.,  2006, \mn@doi [MNRAS]
  {10.1111/j.1365-2966.2006.10316.x}, 369, 425

\bibitem[\protect\citeauthoryear{{Sch{\"a}fer} \& {Heisenberg}}{{Sch{\"a}fer}
  \& {Heisenberg}}{2012}]{2012MNRAS.423.3445S}
{Sch{\"a}fer} B.~M.,  {Heisenberg} L.,  2012, \mn@doi [MNRAS]
  {10.1111/j.1365-2966.2012.21137.x}, \href
  {http://adsabs.harvard.edu/abs/2012MNRAS.423.3445S} {423, 3445}

\bibitem[\protect\citeauthoryear{Sch{\"a}fer, Heisenberg, Kalovidouris  \&
  Bacon}{Sch{\"a}fer et~al.}{2012}]{schafer_validity_2012}
Sch{\"a}fer B.~M.,  Heisenberg L.,  Kalovidouris A.~F.,   Bacon D.~J.,  2012,
  \mn@doi [MNRAS] {10.1111/j.1365-2966.2011.20051.x}, 420, 455

\bibitem[\protect\citeauthoryear{{Schneider} \& {Weiss}}{{Schneider} \&
  {Weiss}}{1988}]{Schneider1988}
{Schneider} P.,  {Weiss} A.,  1988, \mn@doi [\apj] {10.1086/166450}, \href
  {http://adsabs.harvard.edu/abs/1988ApJ...330....1S} {330, 1}

\bibitem[\protect\citeauthoryear{{Seitz} \& {Schneider}}{{Seitz} \&
  {Schneider}}{1994}]{1994A&A...287..349S}
{Seitz} S.,  {Schneider} P.,  1994, \aap, \href
  {http://esoads.eso.org/abs/1994A%26A...287..349S} {287, 349}

\bibitem[\protect\citeauthoryear{Sereno}{Sereno}{2003}]{sereno_gravitational_2003}
Sereno M.,  2003, PRD, 67, 064007

\bibitem[\protect\citeauthoryear{Shapiro \& Cooray}{Shapiro \&
  Cooray}{2006}]{Shapiro:2006em}
Shapiro C.,  Cooray A.,  2006, \mn@doi [JCAP] {10.1088/1475-7516/2006/03/007},
  0603, 007

\bibitem[\protect\citeauthoryear{{Takada} \& {White}}{{Takada} \&
  {White}}{2004}]{2004ApJ...601L...1T}
{Takada} M.,  {White} M.,  2004, \mn@doi [ApJL] {10.1086/381870}, \href
  {http://esoads.eso.org/abs/2004ApJ...601L...1T} {601, L1}

\bibitem[\protect\citeauthoryear{Tansella, Bonvin, Durrer, Ghosh  \&
  Sellentin}{Tansella et~al.}{2017}]{tansella_full-sky_2017}
Tansella V.,  Bonvin C.,  Durrer R.,  Ghosh B.,   Sellentin E.,  2017, ArXiv
  e-prints 1708.00492

\bibitem[\protect\citeauthoryear{Troxel \& Ishak}{Troxel \&
  Ishak}{2015}]{troxel_intrinsic_2015}
Troxel M.~A.,  Ishak M.,  2015, \mn@doi [Physics Reports]
  {10.1016/j.physrep.2014.11.001}, 558, 1

\bibitem[\protect\citeauthoryear{{Van Waerbeke} et~al.,}{{Van Waerbeke}
  et~al.}{2000}]{VanWaerbeke2000}
{Van Waerbeke} L.,  et~al., 2000, \aap, \href
  {http://adsabs.harvard.edu/abs/2000A%26A...358...30V} {358, 30}

\bibitem[\protect\citeauthoryear{{Wang} \& {Steinhardt}}{{Wang} \&
  {Steinhardt}}{1998}]{1998ApJ...508..483W}
{Wang} L.,  {Steinhardt} P.~J.,  1998, \mn@doi [\apj] {10.1086/306436}, \href
  {http://esoads.eso.org/abs/1998ApJ...508..483W} {508, 483}

\makeatother
\end{thebibliography}

\bsp

\appendix
\section{Cross-Correlations}\label{app:1}
We summarise the cross-correlations for all effects considered in \autoref{sect_corrections}, which are expected to be non-zero due to the fact that all originate from the same random field and are thus jointly described by the power spectrum of potential fluctuations. We will use the following abbreviations: Sachs-Wolfe effect (S), integrated Sachs-Wolfe effect (I), temporal Born-corrections (T), peculiar velocity-corrections (V), gravitomagnetic corrections (G) and second order light-propagation (P). All spectra are given in the tomographic bins $i$ and $j$.

\subsection{Redshift space distortions cross-correlations}
Note that we are considering redshift space distortions on the source galaxy redshift. Hence, they correct the gravitational potential in the same way (\ref{eq::rsd}). Furthermore, they all depend on the gravitational field or its parallel derivative, and thus we expect a non-zero cross-correlation between them.

The cross-correlation between the Sachs-Wolfe effect and the integrated Sachs-Wolfe effect gives
\begin{equation}
C^{\mathrm{SI}}_{ij}(\ell) = \frac{2}{c^{7}}\int_0^{\chi_H} \frac{\dd\chi }{\chi^2}\int \frac{\dd^2\ell^\prime}{(2\pi)^2} \left((\ell^\prime)^4 M_{1,\, ij}^{\mathrm{SI}}(\ell^\prime, |\vecdlp| ; \chi)+ (\ell^\prime)^2|\vecdlp|^2 M_{2,\, ij}^{\mathrm{SI}}(\ell^\prime, |\vecdlp| ; \chi)\right),
\end{equation}
where 
\begin{equation}
M^{\mathrm{SI}}_{1,\, ij}(\ell, \ell^\prime ; \chi) =
\int_{\chi}^{\chi_H} \dd \chi^\prime d_i(\chi^\prime)
\int_\chi^{\chi^\prime} \frac{\dd \chi^{\prime\prime}}{(\chi^{\prime\prime})^2}\left(\frac{D_+(a)}{a}\right)_{\chi^{\prime\prime}}\frac{\mathrm{d} }{\mathrm{d} \eta}\left(\frac{D_+(a)}{a}\right)_{\chi^{\prime\prime}} d_j(\chi^{\prime\prime}) P_{\Phi}\left(\frac{\ell}{\chi}; \chi \right)P_{\Phi}\left(\frac{\ell^\prime}{\chi^{\prime\prime}};0\right),
\end{equation}
\begin{equation}
M^{\mathrm{SI}}_{2,\, ij}(\ell, \ell^\prime ; \chi)
=\int_{\chi}^{\chi_H} \dd \chi_1^\prime d_j(\chi^\prime)\int_{0} ^{\chi} \frac{ \dd\chi ^{\prime \prime}}{ (\chi^{\prime\prime})^2}\left(\frac{D_+(a)}{a}\right)_{\chi^{\prime\prime}} \frac{\mathrm{d} }{\mathrm{d} \eta}\left(\frac{D_+(a)}{a}\right)_{\chi^{\prime\prime}} P_{\Phi}\left(\frac{\ell}{\chi}; \chi  \right)P_{\Phi}\left(\frac{\ell^\prime}{\chi^{\prime\prime}};0\right).
\end{equation}
On the other hand, the cross-correlation between the Sachs-Wolfe effect and the peculiar velocities effect is
\begin{equation}
C^{\mathrm{SP}}_{ij}(\ell) = -\frac{1}{c^{5}}\int_0^{\chi_H} \frac{\dd\chi }{\chi^2} \int\frac{\dd k^\prime}{2\pi^2} M_{ij}^{\mathrm{SP}}(k^\prime, \ell; \chi),
\end{equation}
with
\begin{equation}
M^{\mathrm{SP}}_{ij}(k^\prime, \ell; \chi) =
\int_{\chi}^{\chi_H} \dd \chi^\prime d_i(\chi^\prime)\left(\frac{\mathrm{d} D_+(a)}{\mathrm{d}\eta}\right)_{\chi^\prime}
\int_\chi^{\chi_H}\dd \chi^{\prime\prime} \left(\frac{D_+(a)}{a}\right)_{\chi^{\prime\prime}} d_j(\chi^{\prime\prime})k^\prime(\ell^2+(k^\prime\chi^\prime)^2)^2
P_{\Phi\delta}\left(k^\prime;0 \right)
P_{\Phi}\left(\frac{\sqrt{\ell^2+(k^\prime\chi^\prime)^2}}{\chi}; \chi \right)
j_0^\prime(\kp|\chi^\prime-\chi^{\prime\prime}|).
\end{equation}
Finally, the cross-correlation between the integrated Sachs-Wolfe effect and the peculiar velocities effect is 
\begin{equation}
C^{\mathrm{IV}}_{ij}(\ell) = -\frac{2}{c^{6}}\int_0^{\chi_H} \frac{\dd\chi }{\chi^2} \int\frac{\dd k^\prime}{2\pi^2} M_{ij}^{\mathrm{IV}}(k^\prime, \ell; \chi),
\end{equation}
with
\begin{equation}
M^{\mathrm{IV}}_{ij}(k^\prime, \ell; \chi) =
\int_{\chi}^{\chi_H} \dd \chi_1^\prime d_i(\chi_1^\prime) \frac{\dd}{\dd \eta} \left(\frac{D_+(a)}{a}\right)_{\chi_1'} 
\int_{\chi}^{\chi_H} \dd \chi_2^\prime d_j(\chi_2^\prime)\int_0^{\chi^\prime_1}\dd \chi^{\prime\prime} 
\left(\frac{D_+(a)}{a}\right)_{\chi^{\prime\prime}}
k^\prime(\ell^2+(k^\prime\chi_1^\prime)^2)^2 P_{\Phi\delta}\left(k^\prime;0 \right)
P_{\Phi}\left(\frac{\sqrt{\ell^2+(k^\prime \chi_1^\prime)^2}}{\chi}; \chi \right)
j_0^\prime(\kp|\chi_1^\prime-\chi^{\prime\prime}|).
\end{equation}

\subsection{Redshift space distortions cross-correlations with temporal Born-effect.}
The temporal Born-effect correction is very similar to the integrated Sachs-Wolf effect one, since they both depend on the time derivative of the gravitational potential and share a similar structure. Hence, we expect that the temporal Born-correction is correlated with the redshift space distortions ones.   

The cross-correlation between the integrated Sachs-Wolfe effect and the temporal Born-effect is
\begin{equation}
C^{\mathrm{IB}}_{ij}(\ell) = -\frac{4}{c^{9}}\int_0^{\chi_H} \frac{\dd\chi }{\chi^2}\int \frac{\dd^2\ell^\prime}{(2\pi)^2} \left((\ell^\prime)^4 M_{1,\, ij}^{\mathrm{IB}}(\ell^\prime, |\vecdlp| ; \chi)+ (\ell^\prime)^2|\vecdlp|^2 M_{2,\, ij}^{\mathrm{IB}}(\ell^\prime, |\vecdlp| ; \chi)\right),
\end{equation}
with 
\begin{equation}
M^{\mathrm{IB}}_{1,\, ij}(\ell, \ell^\prime ; \chi) =
\frac{\dd}{\dd \eta} \left(\frac{D_+(a)}{a}\right)_{ \chi} \left(\frac{D_+(a)}{a}\right)_\chi \int_{\chi}^{\chi_H} \dd \chi'_1 d_i(\chi'_1)
\int_{\chi}^{\chi'_1}
\frac{ \dd\chi ^{\prime \prime}}{ (\chi^{\prime\prime})^2}\frac{\dd}{\dd \eta} \left(\frac{D_+(a)}{a}\right)_{ \chi''}\left(\frac{D_+(a)}{a}\right)_{\chi^{\prime\prime}} \Phi(\chi^{\prime\prime}) \int_{\chi^{\prime\prime}}^{\chi_H}\dd \chi_2^\prime g_j(\chi,\chi_2^\prime) P_{\Phi}\left(\frac{\ell}{\chi};0 \right)P_{\Phi}\left(\frac{\ell^\prime}{\chi^{\prime\prime}};0\right),
\end{equation}
and
\begin{equation}
M^{\mathrm{IB}}_{2,\, ij}(\ell, \ell^\prime ; \chi) =
\int_{\chi}^{\chi_H} \dd \chi_1^\prime d_i(\chi_1^\prime)
\int_{0}^{\chi}
\frac{ \dd\chi ^{\prime \prime}}{ (\chi^{\prime\prime})^2}\left(\frac{\dd}{\dd \eta} \left(\frac{D_+(a)}{a}\right)_{ \chi''}\right)^2 
\int_\chi^{\chi_H}\dd \chi_2^\prime g_j(\chi,\chi_2^\prime)
P_{\Phi}\left(\frac{\ell}{\chi}; \chi \right)P_{\Phi}\left(\frac{\ell^\prime}{\chi^{\prime\prime}};0\right).
\end{equation}
The Sachs-Wolfe effect and temporal Born-effect cross-correlation is
\begin{equation}
	C^{\mathrm{SB}}_{ij}( \ell) = -\frac{2}{c^8} \int_0^{\chi_H} \frac{\dd \chi}{\chi^2} \int \frac{\dd ^2 \lprime}{(2\pi)^2} (\lprime)^4 M_{ij}^{\mathrm{SB}}(\lprime, |\vecdlp|;\chi),
\end{equation}
with
\begin{equation}
	M^{\mathrm{SB}}_{ij}(\ell, \lprime ; \chi) = \left(\frac{D_+(a)}{a}\right)_\chi\frac{\mathrm{d} }{\mathrm{d} \eta}\left(\frac{D_+(a)}{a}\right)_\chi\int_{\chi}^{\chi_H} \dd \chi^\prime \frac{d_i(\chi^\prime)}{ \chi'^2 } \int_{\chi'}^{\chi_H} \dd \chi^{\prime\prime} g_j(\chi, \chi^{\prime\prime}) P_{\Phi}\left(\frac{\ell}{\chi};0\right)P_{\Phi}\left(\frac{\lprime}{\chi^\prime};\chi'\right).
\end{equation}
Finally, the cross-correlation between the peculiar motions and the temporal Born-effect is
\begin{equation}
	C^{\mathrm{BV}}_{ij}(\ell) = \frac{2}{c^{7}}\int_0^{\chi_H} \frac{\dd\chi }{\chi^2} \int\frac{\dd k^\prime}{2\pi^2} k^{\prime}M_{ij}^{\mathrm{BV}}(k^\prime, \ell; \chi),
\end{equation}
with
\begin{equation}
\begin{split}
M^{\mathrm{BV}}_{ij}(k^\prime, \ell; \chi) &=
\frac{\dd}{\dd \eta} \left(\frac{D_+(a)}{a}\right)_{\chi}\left(\frac{D_+(a)}{a}\right)_{\chi}
\int_{\chi}^{\chi_H} \dd \chi_1^\prime g_i(\chi,\chi_1^\prime)
\int_\chi^{\chi^\prime_1}\dd \chi^{\prime\prime}
\left(\frac{D_+(a)}{a}\right)_{\chi^{\prime\prime}} \\ &
\times  \int_{\chi}^{\chi_H}\dd \chi_2^\prime d_j(\chi_2^\prime)\left(\frac{\dd D_+(a)}{\dd \eta} \right)_{ \chi_2'}
(\ell^2+(k^\prime\chi^{\prime}_2)^2)^2 P_{\Phi\delta}\left(k^\prime;0\right)
P_{\Phi}\left(\frac{\sqrt{\ell^2+(k^\prime \chi^\prime_2)^2}}{\chi};0\right)
j_0^\prime(\kp|\chi_2^\prime-\chi^{\prime\prime}|).
\end{split}
\end{equation}

\subsection{Redshift space distortions cross-correlations with gravitomagnetic effect.}
The gravitomagnetic field is sourced by the momentum density, \eqref{eq::gm_source}, and therefore by potentials and derivatives of the density field. We expect to find a correlation between its correction and the redshift distortions that include angular derivatives of the same potential \eqref{eq::rsd}.

For the Sachs-Wolfe effect we find,
\begin{equation}
	C^{\mathrm{SG}}_{ij}(\ell) = \frac{-2}{c^6} \int_0^{\chi_H} \frac{d \chi}{ \chi^2} \int \frac{dk'}{2 \pi^2} M_{ij}^{\mathrm{SG}}(k', \ell ; \chi ),
\end{equation}
with,
\begin{equation}
	\begin{split}
		M^{\mathrm{SG}}_{ij}(k',\ell; \chi) & = \int_{\chi}^{\chi_H}d \chi' g_i( \chi, \chi')  \frac{\dd}{\dd \eta} \left(\frac{D_+(a)}{a}\right)_{ \chi}   \int_{\chi}^{\chi_H} d \chi'' \left(\frac{D_+(a)}{a}\right)_{ \chi'' } d_j( \chi'') k' ( \ell^2 + ( k' \chi )^2 ) P_{ \delta \Phi}(k';0) P_{\delta\Phi}\left(\frac{ \sqrt{ l^2 + ( k' \chi )^2}}{ \chi}; \chi\right) j_0'( k' | \chi - \chi''|)  
	\end{split}
\end{equation}
On the other hand, the correlation with the integrated Sachs-Wolfe effect
contributes with,
\begin{equation}
	C^{\mathrm{IG}} = \frac{-4}{c^7} \int_0^{\chi_H} \frac{d \chi}{ \chi^2} \int \frac{dk'}{2 \pi^2} M_{ij}^{\mathrm{IG}}(k', \ell ; \chi ),
\end{equation}
where,
\begin{equation}
	\begin{split}
		M^{\mathrm{IG}}_{ij}(k',\ell; \chi ) =  \left( \frac{\dd D_+(a)}{\dd \eta} \right)_{\chi} \int_{\chi}^{\chi_H}d \chi_1' g_i ( \chi , \chi_1') \int_{\chi}^{\chi_H}d \chi_2' d_j( \chi_2') \int_0^{ \chi_2'} d \chi''\frac{\dd}{\dd \eta} \left(\frac{D_+(a)}{a}\right)_{ \chi''} k' P_{\delta\Phi}(k';0) P_{\Phi}\left(\frac{ \sqrt{\ell^2 + (k' \chi )^2}}{\chi}; \chi \right)( \ell^2 + (k' \chi )^2 ) j_0'( k' | \chi - \chi''|)  
	\end{split}
\end{equation}
Finally, peculiar motions add,
\begin{equation}
	C^{\mathrm{VG}}_{ij}(\ell) = \frac{2}{c^5} \int_0^{\chi_H} \frac{d \chi}{ \chi^2} \int \frac{d k'}{2 \pi^2} M_{ij}^{\mathrm{VG}}(k', \ell ; \chi ),
\end{equation}
with,
\begin{equation}
	\begin{split}
		M^{\mathrm{VG}}_{ij}(k', \ell; \chi ) = k'^2 \left( \frac{\dd D_+(a)}{\dd \eta} \right)^2_{\chi}d_i ( \chi ) \int_{\chi}^{\chi_H} d \chi_1' g_i( \chi , \chi_1') \int_{\chi}^{\chi_H} d \chi''  (\ell^2 + (k' \chi )^2 )j_0''(k'| \chi - \chi''|) P_{\Phi}(k';0) P_{\delta\Phi} \left( \frac{ \sqrt{ \ell^2 + (k' \chi )^2}}{\chi}; \chi  \right) 
	\end{split}
\end{equation}
\subsection{Redshift space distortions with second order light propagation}
In this case, there is a correlation between them since both corrections,
\eqref{eq::rsd} and \eqref{eq::velocity},
contain angular derivatives of the potential. The correlation with peculiar velocities,
\begin{equation}
	C^{\mathrm{VP}}_{ij}(\ell) = \frac{2\ell^2}{c^{6}}\int_0^{\chi_H} \frac{\dd\chi }{\chi^2} \int\frac{\dd k^\prime}{2\pi^2} k^\prime M_{ij}^{\mathrm{VP}}(k^\prime, \ell; \chi).
\end{equation}
with
\begin{equation}
M^{\mathrm{VP}}_{ij}(k^\prime, \ell; \chi) =
\left(\frac{D_+(a)}{a}\right)_\chi \int_{\chi}^{\chi_H} \dd \chi^\prime g_i(\chi,\chi^\prime)
\int_\chi^{\chi_H}\dd \chi^{\prime\prime} \left( \frac{\dd D_+(a)}{\dd \eta} \right)_{\chi''} d_j(\chi^{\prime\prime})(\ell^2+(k^{\prime}\chi^{\prime\prime})^2)
P_{\Phi\delta}\left(k^\prime;0 \right)
P_{\Phi}\left(\frac{\sqrt{\ell^2+(k^{\prime}\chi^{\prime\prime})^2}}{\chi}; \chi \right)
j_0^\prime(\kp|\chi-\chi^{\prime\prime}|),
\end{equation}
whereas with the Sachs Wolfe effect,
\begin{equation}
	C_{ij}^{\mathrm{SP}}(\ell) =\frac{2 \ell^2}{c^7} \int_0^{\chi_H} \frac{\dd \chi}{\chi^4} \int \frac{\dd ^2 \lprime}{(2\pi)^2} (\lprime)^2 M_{ij}^{\mathrm{SP}}(\lprime, |\vecdlp|;\chi)
\end{equation}
\begin{equation}
 M^{\mathrm{SP}}_{ij}(\ell, \lprime ; \chi) = \int_{\chi}^{\chi_H} \dd \chi^\prime g_i(\chi,\chi^\prime) d_j(\chi) P_{\Phi}\left(\frac{\ell}{\chi};\chi \right)P_{\Phi}\left(\frac{\lprime}{\chi};\chi\right)
\end{equation}
and finally with the integrated Sachs Wolfe effect,
\begin{equation}
C^{\mathrm{IP}} =-\frac{4 \ell^2}{c^8} \int_0^{\chi_H} \frac{\dd \chi}{\chi^4} \int \frac{\dd ^2 \lprime}{(2\pi)^2} (\lprime)^2 M_{ij}^{\mathrm{IP}}(\lprime, |\vecdlp|;\chi)
\end{equation}
with,
\begin{equation}
 M^{\mathrm{IP}} _{ij}(\ell, \lprime ; \chi) = \int_{\chi}^{\chi_H} \dd \chi^\prime g_i(\chi,\chi^\prime) \int_{\chi}^{\chi_H} \dd \chi^{\prime\prime} d_j(\chi^{\prime\prime}) P_{\Phi}\left(\frac{\ell}{\chi};\chi \right)P_{\Phi}\left(\frac{\lprime}{\chi};\chi\right)
\end{equation}

\subsection{Second order light propagation with temporal Born}
Again, the angular derivatives of the potential in \eqref{eq::rsd} and \eqref{eq::born} produce the following nonzero correlation,
\begin{equation}
	C^{\mathrm{PB}} =\frac{4 \ell^2}{c^9} \int_0^{\chi_H} \frac{\dd \chi}{\chi^4} \int \frac{\dd ^2 \lprime}{(2\pi)^2} (\lprime)^2 M_{ij}^{\mathrm{PB}}(\lprime, |\vecdlp|;\chi)
\end{equation}
with,
\begin{equation}
 M^{\mathrm{PB}}_{ij}(\ell, \lprime ; \chi) = \frac{\dd}{\dd \eta} \left(\frac{D_+(a)}{a}\right)_{\chi} \left(\frac{D_+(a)}{a}\right)_{\chi}\int_{\chi}^{\chi_H} \dd \chi^\prime g_i(\chi,\chi^\prime)
\int_{\chi}^{\chi_H} \dd \chi^{\prime\prime} g_j(\chi,\chi^{\prime\prime}) P_{\Phi}\left(\frac{\ell}{\chi};\chi \right)P_{\Phi}\left(\frac{\lprime}{\chi};0\right)
\end{equation}

\subsection{Second order light propagation with gravitomagnetic effect}
This case is special, since both corrections are evaluated at the same comoving distance $\chi$. While the second-order light propagation correction \eqref{eq::velocity} is always positive due to the gravitational potential being squared, the density currents that source the gravitomagnetic effect \eqref{eq::gm_source} can be positive or negative depending on the neighbouring structure. In an homogeneous universe, we find on average as many negatives as positives currents and therefore the total average vanishes. This is exactly what we obtain numerically, and therefore the correlation between both corrections vanishes,
\begin{equation}
  C_{ij}^{\mathrm{PG}} = 0.
\end{equation}

\subsection{Gravitomagnetic effect with temporal Born-correction}
For these two, the angular derivatives produce as well the following non-zero contribution,
\begin{equation}
	C^{\mathrm{BG}}_{ij}(\ell) = \frac{4}{c ^8} \int_0^{\chi_H} \frac{d \chi}{\chi^2} \int_0^\infty  \frac{d k'}{2 \pi^2} M_{ij}^{\mathrm{BG}}(k', \ell ; \chi ),
\end{equation}
with,
\begin{equation}
	\begin{split}
		M^{\mathrm{BG}}_{ij}(k',\ell; \chi )& = \left( \frac{\dd D_+(a)}{\dd \eta} \right)_{\chi} \left(\frac{D_+(a)}{a}\right)_{\chi}\frac{\dd}{\dd \eta} \left(\frac{D_+(a)}{a}\right)_{\chi} k' \int_{\chi}^{\chi_H}d \chi_1' g_i( \chi , \chi_1') \int_{\chi}^{\chi_H} d \chi_2' g_j( \chi , \chi_2') \int_{\chi}^{ \chi_2'} d \chi'' \left(\frac{D_+(a)}{a}\right)_{\chi''} ( \ell^2 + ( k' \chi )^2 ) j_0'( k' | \chi - \chi''|) \\
																																														  &\times P_{\Phi}(k';0)P_{\delta \Phi} \left( \frac{\sqrt{ \ell^2 + (k' \chi )^2}}{\chi};0 \right).  
	\end{split}
\end{equation}

\label{lastpage}
\end{document}